\documentclass[final,3p,times,twocolumn,authoryear]{elsarticle}

\usepackage{amssymb}
\usepackage{amsmath}
\usepackage{hyperref}
\usepackage[ruled,vlined]{algorithm2e}

\usepackage{booktabs}

\usepackage{makecell}

\usepackage{multirow}

\usepackage{lineno}

\begin{document}

\begin{frontmatter}


\title{A Bayesian multilevel hidden Markov model with Poisson-lognormal emissions for intense longitudinal count data}

\author[inst1]{Sebastian Mildiner-Moraga\corref{cor1}}
\ead{s.mildinermoraga@uu.nl}
\cortext[cor1]{corresponding author.}
\author[inst1]{Emmeke Aarts}

\affiliation[inst1]{organization={Department of Methodology and Statistics, Utrecht University},
            addressline={Padualaan 14}, 
            postcode={3584 CH}, 
            city={Utrecht},
            country={the Netherlands}}

\begin{abstract}
Hidden Markov models (HMMs) are probabilistic methods in which observations are seen as realizations of a latent Markov process with discrete states that switch over time. Moving beyond standard statistical tests, HMMs offer a statistical environment to optimally exploit the information present in multivariate time series, uncovering the latent dynamics that rule them. Here, we extend the Poisson HMM to the multilevel framework, accommodating variability between individuals with continuously distributed individual random effects following a lognormal distribution, and describe how to estimate the model in a fully parametric Bayesian framework.  The proposed multilevel HMM enables probabilistic decoding of hidden state sequences from multivariate count time-series based on individual-specific parameters, and offers a framework to quantificate between-individual variability formally.  Through a Monte Carlo study we show that the multilevel HMM outperforms the HMM for scenarios involving heterogeneity between individuals, demonstrating improved decoding accuracy and estimation performance of parameters of the emission distribution, and performs equally well when not between heterogeneity is present. Finally, we illustrate how to use our model to explore the latent dynamics governing complex multivariate count data in an empirical application concerning pilot whale diving behaviour in the wild, and how to identify neural states from multi-electrode recordings of motor neural cortex activity in a macaque monkey in an experimental set up. We make the multilevel HMM introduced in this study publicly available in the R-package \verb|mHMMbayes| in CRAN.
\end{abstract}




\end{frontmatter}


\section{Introduction}
\label{sec:general_introduction}

Longitudinal count data are commonly encountered in fields such as healthcare \citep{Inaba2017, Chiang2018, conesa_bayesian_2015, marchuk_predicting_2018}, ecology \citep{deruiter_multivariate_2017,perry_hidden_2019}, neurophysiology \citep{radons_analysis_1994, van_kempen_top-down_2021, kirchherr_bayesian_2022}, seismology \citep{orfanogiannaki_identifying_2010, felix_poisson_2022}, meteorology \citep{Holsclaw2017}, traffic flow and crash models \citep{cheng_comparison_2017, rossi_poisson_2015}, and finance, among others \citep[for a review, see,][]{davis_count_2021}. Analyzing count time series requires dealing with the autocorrelation of the data and possibility of over-dispersion. Hidden Markov models (HMMs) have been widely used to model such data, which assume that the observations are generated by an unobserved Markov chain. HMMs are flexible probabilistic models for sequential data which assume the observations to depend on an underlying latent state process. They are used for classification tasks, forecasting, or general inference on the data-generating process \citep[for an overview, see,][]{Zucchini2017}. In an HMM's basic model formulation, the underlying state sequence is assumed to be a finite-state first-order Markov chain. This assumption is mathematically and computationally very convenient and allows for an efficient likelihood evaluation and inference \citep{Zucchini2017} and allows to deal with the autocorrelation in the longitudinal series of observations. When it comes to count longitudinal data, the HMM-Poisson is a popular variant, which assumes that the counts are generated by a Poisson emission distribution with mean that depends on the underlying state. Several applications of the HMM-Poisson are found in the literature: for example, states modelling levels of risk of epileptic seizures \citep{Chiang2018} and advancement of pulmonary disease \citep{marchuk_predicting_2018}, inferring four different behaviours in pilot whales using sensor data \citep{deruiter_multivariate_2017}, uncovering neural states involved in the different stages of the orchestration of a motor behaviour \citep{sadacca_behavioral_2016}, or early warning detection of influenza outbreaks \citep{conesa_bayesian_2015}. However, the HMM-Poisson does not account for heterogeneity between individuals or clusters of data.

In practice, it is common for longitudinal count data to exhibit heterogeneity between individuals or groups, which can lead to biased and inaccurate estimates if not accounted for and worsen the effect of over-dispersion in the count data \citep{mcneish_poisson_2019}. Multilevel models have been developed to address this issue, where individual-specific parameters are assumed to follow a population-level distribution \citep{hox_multilevel_2018}, including multilevel HMMs \citep{Altman2007}. Although several methods have been proposed to model between individual variability, the estimation of individual specific random effects on both components of the model via traditional maximal likelihood methods tends to be computationally intractable using frequentist approaches, possibly limiting a more widespread utilization of these models \citep{Altman2007, mcclintock_worth_2021}. For this reason most of the application of the multilevel HMM consist on mixture models that allocate variability between groups of individuals with discrete random effects, limiting the possibility of exploring individual differences \citep[e.g.,][]{maruotti_multilevel_2021}. However, recent research has shown that wrongly assuming a mixture distribution for a continuously distributed between individual variation can lead to biased parameter estimation \citep{mcclintock_worth_2021}. A small number of applications have shown that it is possible to include individual specific random effects in both components of HMMs with more than two hidden states using an iterative Bayesian estimation framework \citep[e.g.,][]{Shirley2012a, DeHaan-Rietdijk2017, vidal_bustamante_fluctuations_2022, hale_hidden_2023, mildiner_moraga_evidence_2023}. Yet, to the best of our knowledge, only \citet{desantis_hidden_2011} and \citet{kirchherr_bayesian_2022} considered a multilevel HMM with continuous random effects in combination with an emission distribution suitable to analyze count data. However, the multilevel HMM with zero inflated Poisson conditional-probabilities in \citet{desantis_hidden_2011} only included individual-specific random effects on the emission distribution and was limited to two hidden states due to estimation constraints. On the other hand, we developed the model in \citet{kirchherr_bayesian_2022} in collaboration with applied researchers, which we formally introduce, describe, and assess here for a statistical audience.
 
The purpose of this study is to introduce a novel Bayesian multilevel hidden Markov model with a Poisson-lognormal emission distribution (MHMM-PLN) to model longitudinal count data of multiple individuals at a time. Our proposed model extends the standard HMM to the multilevel framework to account for the hierarchical structure of the data and incorporates a lognormal distributed individual specific random effects that captures the over-dispersion between individuals in the count data, and allows to study individual differences between them. We describe how to efficiently sample the group-level and individual specific parameters in the model in the Bayesian framework, which we made openly accessible on the R library \verb|mHMMbayes| \citep{aarts_mhmmbayes_2019}. We show that the model has the potential to provide more accurate estimates of the sequence underlying states and emission parameters than a comparable single-level HMM under data generating conditions with or without between individual heterogeneity in one or both components of the model with a small Monte Carlo simulation. 

The rest of the manuscript is organized as follows. Section \ref{sec:model_specification} describes the specification and estimation of a basic HMM Poisson and the multilevel HMM Poisson-lognormal model. Section \ref{sec:sim_study} consists of the Monte Carlo simulation study aimed at evaluating the mulilevel HMM. In Section \ref{sec:emp_applications} we showcase the information that can be obtained with the model with two distinct empirical applications. We conclude with a discussion of the main results, limitations, and future perspectives in Section \ref{sec:discussion}.

\section{Model specification}
\label{sec:model_specification}

In this section we present a brief introduction to the Bayesian HMM for count time series and describe approaches to apply it to a group of individuals. Subsequently, we describe the Bayesian multilevel HMM for count time series and discuss how it extends the modeling capabilities of the single-level HMM. Finally, we discuss inference of the MHMM using Bayesian estimation.

\subsection{Specification of the single-level HMM Poisson-lognormal}
\label{sec:basic_hmm}

\begin{figure*}[h]
\centering
\includegraphics[width=0.7\linewidth]{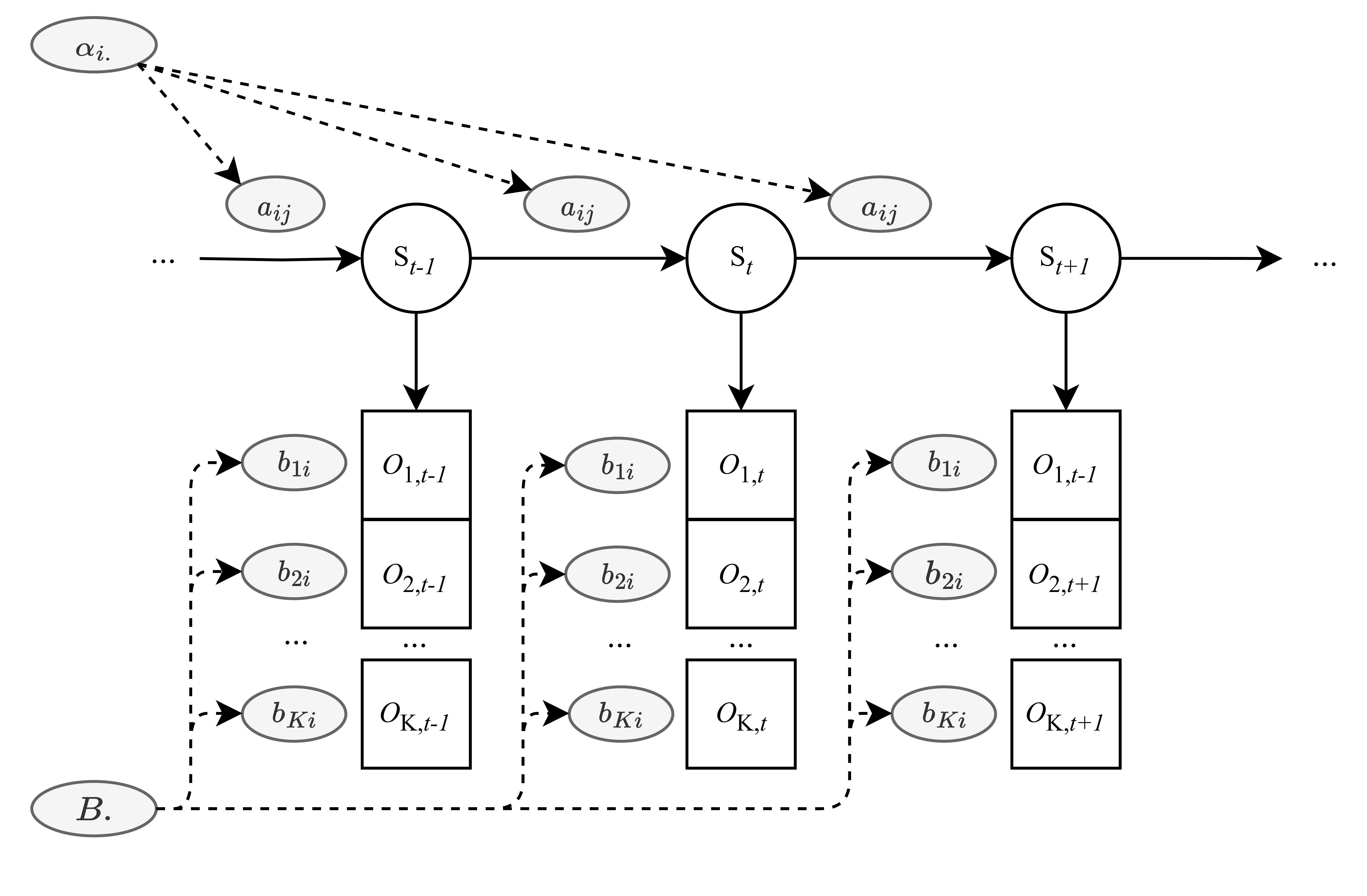}
\caption{Directed acyclic graph of a basic hidden Markov model with a multivariate Poisson emission distribution. The probability $a_{ij}$ of switching to a hidden state $j$ at time $t$, depicted as $S_t = j$ with a circle, depends only on the state $i$ at the previous time point (i.e., $S_{t-1} = i$). The multivariate counts observed at time $t$ over the $K$ time series, depicted with squares as $O_{1t}, ..., O_{Kt}$, depends only on the value of the current latent state $S_t$. Superimposed are the model parameters, with transition probabilities $a_{ij}$, Poisson means $b_{1i}, ..., b_{Ki}$, the multinomial logit intercepts $\alpha_{i.}$ and the emission matrix $B_{.}$, respectively.}
\label{fig:dag_hmm_pln}
\end{figure*}

In the basic single-level Hidden Markov Model (HMM), we assume that one or more time series of count observations, denoted as $O_1, O_2, ..., O_T$ across time points $t \in \{1, 2, ..., T\}$, result from a sequence of underlying states, represented by $S_1, S_2, ..., S_T$, where there are $M$ possible hidden states $S_t \in \{1,...,M\}$, which cannot be directly measured. More formally, the model posits that the probability of observing a count $O_t = q$ at time $t$ depends solely on the state $S_t$ visited at that time, denoted as $P(O_t=q | S_t=i)$, and is defined by an arbitrary emission distribution. The hidden states are not identically and independently distributed; instead, they follow a first-order Markov process. In other words, the probability of transitioning from state $S_{t-1} = i$ at time point $t-1$ to state $S_{t} = j$ at time $t$ depends only on the departing state $i$ at time point $t$, expressed as $P(S_t = j | S_{t-1} = i)$. Figure \ref{fig:dag_hmm_pln} provides a detailed illustration of the general structure of a basic HMM for multivariate count data.

Three sets of parameters describe the basic single-level HMM: the initial distribution of the states $\pi$ with the initial probabilities $\pi_i \equiv P(S_{t=1}=i)$ of each state at the first-time measurement (i.e., $t=1$); the transition distribution $A$ containing the probabilities $a_{ij} = P(S_t = j | S_{t-1} = i)$ of switching between states; and the emission distribution $B$ consisting on the parameters $b_i$ defining $P(O_t = q | S_t=i)$, the probabilities of observing $q$ counts each time series, given the hidden state $S_t$ being visited.

In terms of these three components, the likelihood $L$ of a basic univariate hidden Markov model (HMM) can be defined as

\begin{equation}
    L = \pi B(O_1) AB(O_2) AB(O_3) ... AB(O_T)1' 	
\end{equation}

where $\pi$ is a column vector of dimensions $(1 \times M)$ with the initial probabilities of the states $S_{t=1} \in \{1,...,M\}$, $\pi_i \equiv P(S_{t=1} = i)$, $A$ is a transition probability matrix of dimensions $(M \times M)$ denoting the transition probabilities $a_{ij} = P(S_t = j | S_{t-1} = i)$ which is assumed to be homogeneous over the occasions $t$, $B$ is the emission distribution consisting of a diagonal matrix of dimensions $(M \times M)$ with the \textit{ith} diagonal element denoting the emission probability density $b_i$ defining $P(O_t = q | S_t=i)$ which can take any arbitrary form, and $1'$ is a column vector of $M$ elements of value one. In the context of this study, a Poisson emission distribution is assumed for the observed  counts $O_t$, given the hidden state $S_t$, such that $P(O_t=q | S_t=i) \equiv \text{Poisson} (b_i) $. $B$ is the diagonal matrix of dimensions $(M \times M)$ with its diagonal elements denoting the Poisson parameters $b$ (expected  counts) for the $M$ hidden states.

Although the model has been defined for an univariate observed sequence, it can readily be extended to multivariate data with the convenient assumption that the multiple sequences (e.g., observations from multi-observed sequence probes) are conditionally independent given the hidden states \citep[e.g.,][see diagram in Figure \ref{fig:dag_hmm_pln}]{Zucchini2017}. That is, we specify the joint distribution $P(O_{1t} = q_1, O_{2t}  = q_2, ..., O_{kT} = q_T)$ for the  $k \in \{1,2,...,K\}$ state-dependent emission distributions as the product of the $K$ marginal state-dependent emission probability densities $P(O_{kt} = q_t | S_t=i)$:

\begin{equation}
\begin{split}
        P(O_{11} = q_1, ..., O_{1T} = q_T) ... P(O_{K1} = q_1, ..., O_{KT}  = q_T) = \\
        = \prod_{t=1}^T \prod_{k=1}^K P(O_{kt} = q_t | S_t=i) 
\end{split}
\end{equation}

As a result, the likelihood of the general multivariate HMM takes the form:

\begin{equation}
    L = \prod_{k=1}^K \pi B_k(O_{k1}) AB_k(O_{k2})AB_k(O_{k3}) ...AB_k (O_{kT})1'
\end{equation}

A basic HMM of this specification can be easily estimated using direct likelihood maximization \citep[e.g.,][]{Zucchini2017}, following a computationally efficient recursive implementation of the Expectation-Maximization algorithm \citep{dempster_maximum_1977} known as the Baum-Welch algorithm \citep{baum_maximization_1970}, or via the Bayesian estimation framework as we do here \citep[e.g.,][]{Scott2002}.

The Bayesian specification requires choosing a suitable set of priors for the transition and emission parameters. Although a categorical distribution and its conjugate Dirichlet prior could be used for the transition probabilities $a_{ij}$, for consistency with the specification of the multilevel HMM in Section \ref{sec:multilevel_hmm} here we adopt a multinomial logit with a normal prior of the form:

\begin{equation}
    a_{ij} \sim MNL(\alpha_{ij})
\end{equation}
\begin{equation}
    \alpha_{ij} \sim N(\alpha_{0ij}, \Psi_{0ij}  )
\end{equation}

where $\alpha_{0ij}$ represent the prior expectations on the intercepts of the multinomial logit, and $\Psi_{0}$ the prior expectation on their covariance.

The multinomial logistic regression is a common way to model categorical outcomes, such as the choice of one of several possible states, where the probabilities for selecting outcomes can be explained by covariates. Hence, probabilities to transition from hidden state $i \in \{1,..,M\}$ to state $j \in \{1,..,M\}$ is modeled using $M$ batches of $M-1$ intercepts $\alpha_{ij}$. That is,

\begin{equation}
    a_{ij} = \frac{exp(\alpha_{ij})}{1 + \sum_{l=2}^M exp(\alpha_{il})}
\end{equation}

where the numerator is set equal to $1$ for $j = 1$, making the first state of every row of the transition probability matrix $A$ the baseline category. Notice that the multinomial logistic regression consists on a matrix of $(M \times M-1)$ elements, with M rows denoting transitions from $M$ states, and $M-1$ columns denoting transitions to $M-1$ states, since the transitions to the first state (first column) are defined as the reference category and are not estimated (thus, $\alpha_{ij=i}=0$).

We let the state $i$--specific Poisson parameters $b_{ki}$ for each of the $K$ dependent variables follow a lognormal distribution of the form: 

\begin{equation}
    ln(b_{ki}) \sim N(b_{0ki}, \tau_{0ki})
\end{equation}

where $b_{0ki}$ and $\tau_{0ki}$ represent the prior expectations on log--transform means and uncertainty of the Poisson parameters $b_{ki}$, respectively. Note that using a normal prior on the log-transform of the Poisson parameters $b_{nki}$ is equivalent to specifying a lognormal prior on the untransformed Poisson parameter. In addition, as previously mentioned, because $b_{ki}$ are assumed to be conditionally independent given the state, no covariance matrix is estimated for the covariance between dependent variables. 

\subsection{Dealing with data of multiple individuals}
\label{sec:multi_ind_hmm}

When applying Hidden Markov Models (HMM) to multiple sequences of data, researchers face challenges in handling individual variation. One approach involves fitting a "complete pooling" standard HMM, assuming a same set of parameters shared among all individuals --with the option of incorporating individual covariates, making parameters conditional on these attributes. However, complete pooling does not account for individual variation and may lead to biased inference \citep{Rueda2013}. Conversely, fitting a separate "unpooled" HMM for each individual (i.e., "no pooling") captures individual heterogeneity but is inefficient and may lead to difficulties in comparing parameters and hidden states between individuals. The individual fixed effects model \citep[e.g.,][]{jonsen_joint_2016, mcclintock_momentuhmm_2018} combines aspects of both approaches --it assumes "complete pooling" for one of the components of the model, and "no pooling" to the other-- but results in complex and less interpretable models. An intermediate solution is the mixture HMM \citep[e.g.,][]{bartolucci_discrete_2015, maruotti_multilevel_2021, langrock_flexible_2012}, where latent classes with varying model parameters are specified, while individuals within each class share common parameters. However, it assumes a limited number of homogeneous subgroups and may be a poor approximation when individual variation is continuous, as shown by \citet{mcclintock_worth_2021}.

The continuous random effects multilevel framework (i.e., "partial pooling"), on the other hand, brings together the benefits of both "complete pooling" and "unpooled" models: it allows estimation of parameters at both the individual and group levels, providing group-level estimates while allowing deviation between individual parameters, and measuring individual variation \citep{gelman_beyond_2014}. Expanding the model into the multilevel framework with continuous random effects, the MHMM presents three main advantages compared to traditional HMMs: 1) it estimates individual parameters that deviate from group means according to a parametric continuous distribution (e.g., normal), making possible to decode distinct underlying processes over time for each of them, while maintaining consistent meanings for each state across individuals, 2) it enables the measurement of deviations from the group-level means (i.e., between-individual heterogeneity) through those random effects, and 3) it incorporates a hierarchical structure that imparts regularization to individual-specific parameters, enhancing the model's resilience to outliers; all this while using all the information available in the data.

In the upcoming subsection, we describe a statistical specification in the Bayesian framework of a MHMM with a Poisson-lognormal emission distribution suitable to count time series.

\subsection{Specification of the multilevel HMM Poisson-lognormal}
\label{sec:multilevel_hmm}

\begin{figure*}[h]
\centering
\includegraphics[width=0.7\linewidth]{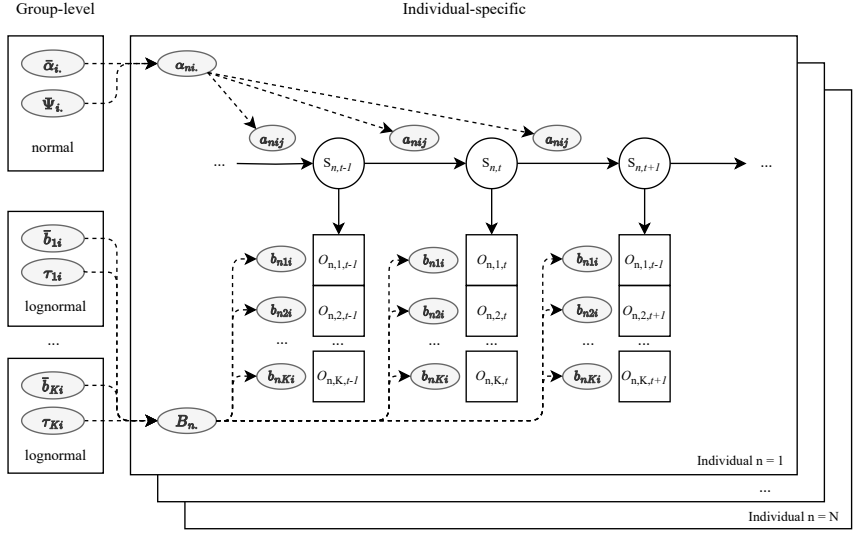}
\caption{Directed acyclic graph of a multilevel hidden Markov model with a multivariate emission distribution. Each hidden state Sn for individual $n \in \{1, 2, ..., N\}$ over the time points $t \in \{1, 2, ..., T\}$, depicted with circles, depends only on the state at the previous time point. The observed data $O_{n,k,1}, ..., O_{n,k,T}$ for outcome variables $k \in \{1, 2, ..., K\}$, depicted with squares, depends only on the value of the current latent state $S_{n,t}$. Superimposed are the model parameters, with individual n specific transition probabilities $a_{nij}$ and  Poisson emission means $b_{nki}$, the individual specific multinomial logit intercepts $\alpha_{ni.}$, and the emission matrix $B_{n1}$. Also depicted are the group-level mean $\bar{\alpha}_{i.}$ and covariance $\Psi_{i.}$ of the normal distribution on $\alpha_{ni.}$, and the group-level means $\bar{b}_{1i}, ..., \bar{b}_{Ki}$ and variances $\tau_{1i}, ..., \tau_{Ki}$ of the normal distribution on $b_{1Ki}, ..., b_{nKi}$.}
\label{fig:dag_mhmm_pln}
\end{figure*}

In the current implementation of the multilevel HMM (see Figure \ref{fig:dag_mhmm_pln}) we again use a multinomial logistic regression to represent $a_{nij}$, the probability for an individual $n \in \{1,..,N\}$ of switching between a state $S_{n,t-1} = i$ and state $S_{n,t} = j$, at occasion $t$

\begin{equation}
    a_{nij} \sim MNL(\alpha_{nij})
\end{equation}

We adopt a group-level state-specific multivariate normal prior distribution on the individual-specific parameters $\alpha_{nij}$, with mean vector $\bar{\alpha}_{i.}$ of $M-1$ elements denoting the group-level intercepts $\bar{\alpha}_{ij}$, and covariance $\Psi_{i.}$ denoting the covariance between the $M-1$ state $i$-specific intercepts over individuals. A convenient hyper-prior on the parameters of the group-level prior distribution is a multivariate normal distribution for the mean vector $\bar{\alpha}_{i.}$, and an Inverse Wishart distribution for the covariance $\Psi_{i.}$. That is,

\begin{equation}
    \alpha_{nij} \sim N(\bar{\alpha}_{ij}, \Psi_{ij}  )
\end{equation}
\begin{equation}
    \bar{\alpha}_{ij} \sim N(m_{0ij}, 1/K_0 \Psi_{ij}  )
\end{equation}
\begin{equation}
    \Psi_{ij} \sim IW(\Psi_{0}, df_0)
\end{equation}

The parameters $m_{0ij}$, and $K_0$ denote the values of the parameters of the hyper-prior on the group (mean) vector $\bar{\alpha}_{ij}$. Here, $m_{0ij}$ represent the prior expectations on the group-level transition intercept means, and $K_0$ denotes the number of observations (i.e., the number of hypothetical prior individuals) on which $m_{0ij}$ are based. The parameters $\Psi_{0}$ and $df_0$, respectively, denote values of the covariance and the degrees of freedom of the hyper-prior for the Inverse Wishart distribution on the group variance $\Psi_{ij}$ of the individual-specific random intercepts $\alpha_{nij}$.

Just as for the basic HMM introduced in section \ref{sec:basic_hmm}, count data $O_{nkt}$ follow Poisson emission distributions, but here with individual-specific parameters $b_{nki}$ conditionally independent on the state $S_{t} \in \{1,..,M\}$ (Figure \ref{fig:dag_mhmm_pln}). That is,

\begin{equation}
    O_{nkt} \sim Poisson(b_{nki})
\end{equation}

for observed sequence (the dependent variable) $k \in \{1, ..., K\}$, on time point $t \in \{1, ..., T\}$, and individual $n \in \{1, ..., N\}$, given the state $i$.

We adopt a group-level state-specific lognormal prior distribution on the individual-specific parameters $b_{nki}$, with group-level means $\bar{b}_{ki}$, and variancees $\tau_{ki}$. The complete specification of the emission distribution adopted is detailed in probabilistic notation below, along with conveniently defined hyper-priors on the parameters of the model

\begin{equation}
    log(b_{nki}) \sim N(\bar{b}_{ki}, \tau_{ki})
\end{equation}
\begin{equation}
    \bar{b}_{ki} \sim N(l_{0ki}, \tau_{0ki})
\end{equation}
\begin{equation}
    \tau_{ki} \sim IG(c_{ki}, d_{ki})
\end{equation}

The parameters $l_{0ki}$ denote the \textit{a priori} expectations on the group-level mean counts on the logarithmic scale along with the \textit{a priori} expectations on their variability, $\tau_{0ki}$, respectively. Finally, $c_{ki}$, $d_{ki}$ are the hyper parameters denoting the shape and rate of the hyper prior on the between-individual variance $\tau_{ki}$. Note that using a normal prior on the log-transformation of the individual-specific, observed sequence-specific, state-specific Poisson parameters $b_{nki}$ is equivalent to specifying a lognormal prior on the untransformed Poisson parameter. In addition, as previously mentioned, because $b_{nki}$ are assumed to be conditionally independent given the state, no covariance matrix is estimated for the covariance between dependent variables. 

In \verb|mHMMbayes|, models are fitted using a hybrid Metropolis within Gibbs MCMC algorithm, which expands on classic HMM implementations by using Bayesian estimation as outlined in \cite{Scott2002} (see Algorithm \ref{mhmm_estimation_algo} in Supplementary materials). Thus, we follow a MCMC sampler algorithm to iteratively sample from the appropriate conditional posterior distributions of ($\bar{\alpha}_{ij}$, $\bar{b}_{ki}$, $\Psi_{ij}$, $\tau_{ki}$) given the remaining parameters in the model. When natural conjugate priors are available, Gibbs estimation algorithm is used due to its simplicity and computational efficiency. However, since the normal and the lognormal group-level priors respectively chosen for $\alpha_{nij}$ and $b_{nki}$ are not conjugate, we use a self-scaled Random Walk Metropolis Hastings sampling procedure to approximate their posterior distribution. Without entering into too much detail, on each new iteration it proposes new candidate values for the individual-specific posterior distributions, drawing a random step from a univariate or multivariate normal distribution whose (co-)variance the covariance matrix is set to be a combination of the covariance matrix obtained from the individual data and the group-level covariance matrix \citep[for more details, refer to][]{rossi_hierarchical_2005}.

\section{Simulation study}
\label{sec:sim_study}

In this section, we introduce and conduct a Monte Carlo simulation study to evaluate the performance of the multilevel HMM ("partial pooling" model), and additionally, compare it to the performance of a single-level HMM ("complete pooling" model) for data with and without (continually distributed) individual variability in the parameters of none, one, or both the transition and emission distributions.

To keep the simulation consistent with real prospective applications, group-level parameters used in to simulate data are based on the empirical results of fitting a single-level HMM with a Poisson emission distribution of a single dependent variable, the number of breath anomalies per 15 minutes bin, in 51 patients found in \cite{marchuk_predicting_2018}. Data for $N = 50$ individuals with $T = 500$ observations\footnote{Preliminary results showed that using $T=5000$, more in line with T in \cite{marchuk_predicting_2018}, only improved the parameter estimation marginally with a large increment in its computational cost.} each were generated for $N_{rep}=250$ repetitions fixing the number of hidden states to $M = 4$, with a group-level transition probability matrix\footnote{Same as per \cite{marchuk_predicting_2018}, probabilities $<0.005$ are rounded to $0.00$.} (group-level transition means) $\bar{A} = \biggl(\begin{smallmatrix}
  0.85 & 0.13 & 0.02 & 0.00 \\
  0.23 & 0.63 & 0.13 & 0.01\\
  0.07 & 0.24 & 0.63 & 0.06\\
  0.03 & 0.04 & 0.14 & 0.79
\end{smallmatrix}\biggl)$, group-level Poisson emission parameters (group-level Poisson-lognormal emission means) $\bar{B} = log(1, 11, 38, 119)$, and a between-individual variability according to the corresponding simulation condition (see below). Although the model allows for the incorporation of time-invariant individual-specific covariates in any of the two components of the model, none were included as part of the data generation procedure.

\subsection{Simulation conditions}

Data was simulated under four levels of individual heterogeneity: (i) no heterogeneity between individuals in the transition and emission means ($\Psi_{i.} = \text{diag($0, 0, 0, 0$)})$, $\tau_{.} = (0, 0, 0, 0)$), (ii) only individual heterogeneity in the transition means ($\Psi_{i.} = \text{diag($0.9, 0.9, 0.9, 0.9$)})$, $\tau_{.} = (0,0,0,0)$), (iii) only individual heterogeneity in the emission Poisson means ($\Psi_{i.} = \text{diag($0, 0, 0, 0$)})$, $\tau_{.} = (0.9, 0.7, 0.5, 0.2)$), and (iv) individual heterogeneity in both the transition and emission means ($\Psi_{i.} = \text{diag($0.9, 0.9, 0.9, 0.9$)})$, $\tau_{.} = (0.9, 0.7, 0.5, 0.2)$). Notice that the values chosen for the between-individual variance components $\Psi$ and $\tau$ are well in line with the amount of variability reported in the literature for empirical applications of related mixed HMMs (e.g., 0.83 in \citealt{Jackson2015b}; 0.10, 0.50 in \citealt{maruotti_semiparametric_2009}; 0.20, 0.42 in \citealt{mcclintock_worth_2021}; 0.10-1.00 in \citealt{Rueda2013}; 1.0 in \citealt{zhang_study_2014}), with $\Psi_{i.} = \text{diag($0.9, 0.9, 0.9, 0.9$)})$ corresponding to standard deviations in the range $(0.007-0.225)$ for the between-individual heterogeneity on the probability scale, and $\tau_{.} = (0.9, 0.7, 0.5, 0.2)$ resulting in standard deviations of $(1.89, 15.72, 39.30, 61.88)$ between individual-specific Poisson parameters $b_{n.}$.

\subsection{Model fitting}
\label{sec:model_fitting}

Data simulation and model fitting was performed using R \citep{RCoreTeam2020} and the developer version of the package \verb|mHMMbayes| \citep{aarts_mhmmbayes_2019}. The model was fit with 4000 iterations, with the first 2000 being discarded to dissipate the effect of starting conditions (burn-in). In fitting the model, agnostic non-informative hyper-priors were specified for all group-level parameters. Convergence of all group-level parameters was checked by visual inspection of three randomly drawn simulation repetitions per scenario, for which an additional chain with randomized starting values was used. The sequence of most likely states given the observations was determined with the Viterbi algorithm \citep{Forney1973, Viterbi1967} based on individual-specific parameters.

\subsection{Evaluation}

Model performance was evaluated based on mean and relative mean bias, precision (empirical standard error), and coverage of the 95\% credibile interval (CrI) for the group-level parameters $\bar{\alpha}_{ij}$ and $\bar{b}_{ij}$, and the individual random effects $\Psi_{ij}$ and $\tau_{ki}$ on both components in the model. Model performance of the group-level transition parameters $\bar{\alpha}_{ij}$ is presented on the probability domain ($\bar{a}_{ij}$) to aid interpretation. Note however that parameter estimation is on the logit domain. Model performance of the variance of the individual level random effects $\Psi_{ij}$ and $\tau_{ki}$ are presented on the logit domain.
State decoding accuracy by the Viterbi algorithm \citep{Viterbi1967}, also known as global state decoding, was evaluated in terms of balanced accuracy, F1 score, and Cohen's $\kappa$.

For the code to reproduce the Monte Carlo simulation, refer to \url{https://doi.org/10.5281/zenodo.10834837}.

\subsection{Simulation results}
\label{sec:sim_results}

Visual inspection of the parallel chains revealed satisfactory convergence of all the group-level fixed effects in the two models (i.e., convergence rate 100\%), independently from the condition of between-individual heterogeneity specified. The convergence of transition random effects was more challenging, with an average of 92\% of the parameters assessed satisfactorily meeting the criterion. More specifically, the average convergence rate of transition random effects were 100\% in scenario 1 (no between-heterogeneity), 86\% in scenario 2 (heterogeneous transitions), 100\% in scenario 3 (heterogeneous emissions), and 81\% in scenario 4 (heterogeneity in both components). All but one of the emission random effects (98\%) showed satisfactory convergence, with one of the random effects exhibiting divergence between the chains in a scenario with between-heterogeneity in both components of the model.

\subsubsection{Transitions}

\begin{figure*}[ht]
\centering
\includegraphics[width=0.7\linewidth]{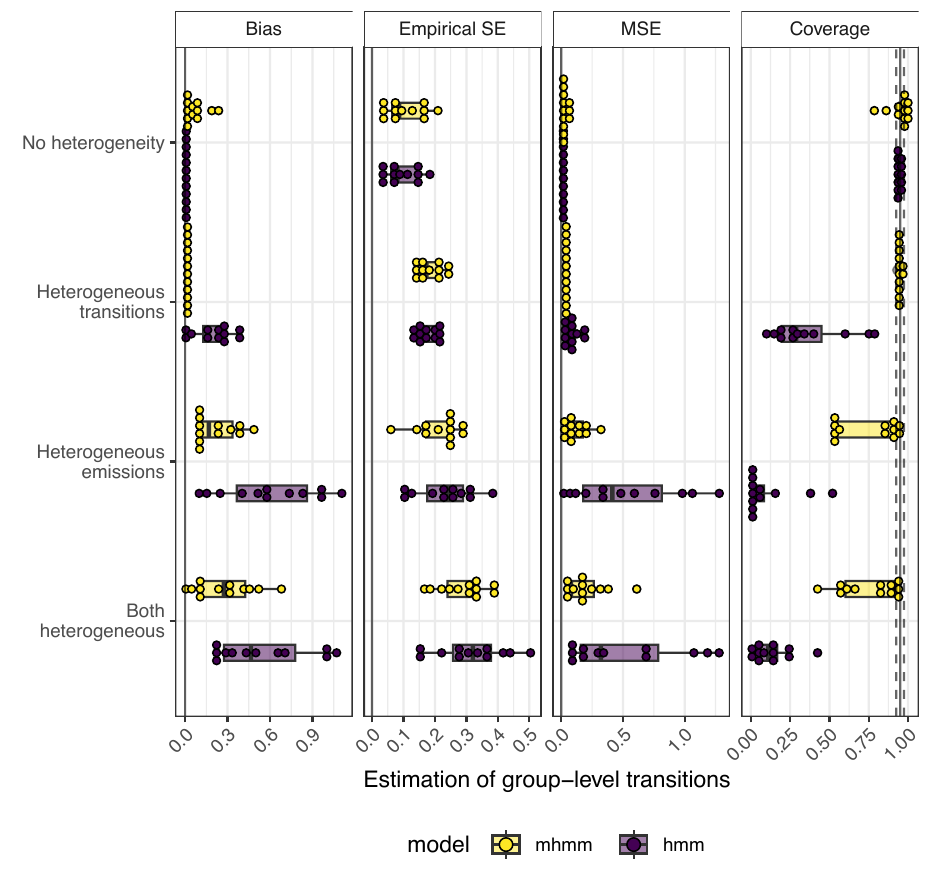}
\caption{Estimation performance group-level transitions transition means $\bar{\alpha}_{ij}$ on the MHMM (in yellow) and the HMM (in purple) over the four conditions of between-individual heterogeneity (no heterogeneity, heterogeneous transitions, heterogeneous emissions, both heterogeneous). Dots represent \textit{maximum a posteriori} parameter point-estimates, with overlaying box-plots to facilitate interpretation.}
\label{fig:gamma_fixed}
\end{figure*}

\textit{Group-level means}. We compared the estimation performance of transition probabilities between the multilevel HMM and the HMM over the four conditions of between individual heterogeneity. The results are visualized in Figure \ref{fig:gamma_fixed} (for exact values in all parameters, see Tables \ref{table:supp_1}--\ref{table:supp_4}). Overall, our results indicate that when between individual heterogeneity is present, the MHMM performs better than the HMM in terms of bias, empirical SE, MSE, and parameter coverage. Under conditions of no between individual heterogeneity the performance gap between the two models became less substantial.

In Scenario 1 (Table \ref{table:supp_1}), where no between-heterogeneity was present in either component of the model, both models exhibited good estimation performance. No substantial biases were observed in estimating group-level transition means, with absolute bias of $<0.001$ and $<0.005$ for the HMM and the MHMM, respectively. Both models presented an empirical standard error under $0.011$ and an MSE lower than $0.0001$. However, the coverage of the 95\% credible intervals was more accurate in the HMM (mean $=.945$, range: $.917$-$.980$) compared to the MHMM, which presented over-coverage (mean $=.964$, range: $.794$-$1.000$).

In Scenario 2 (Table \ref{table:supp_2}), where only the transition distribution is heterogeneous between individuals, the MHMM outperformed the single-level HMM. The MHMM accurately estimated all transition probabilities group-level parameters with a relative bias under the $\pm10\%$ threshold, while the HMM  exhibited a relative bias under the $\pm10\%$ threshold for only half of the group-level parameters. Parameter estimation in the MHMM was characterised by lower bias (mean absolute bias over transitions parameters of $0.001$ compared to $0.017$ for the HMM), comparable precision (mean empirical standard error of $0.017$ and $0.016$ for the MHMM and the HMM, respectively), and substantially improved coverege of the 95\% credible intervals (mean coverage of $.944$ with range: $.924$-$.956$ for the MHMM, compared to $.280$ with range: $.108$-$.572$ for the HMM). Even after bias correction, the 95\% credible intervals remained overconfident in the HMM, leading to under-coverage.

In both Scenario 3 (Table \ref{table:supp_3}), where between-heterogeneity was present only in the emission distribution, and Scenario 4 (Table \ref{table:supp_4}), where between-heterogeneity was present in both components, the MHMM consistently outperformed the single-level HMM. In Scenario 3, the MHMM presented a lower mean absolute bias (MHMM: $0.020$, HMM $0.033$), lower mean absolute relative bias ($0.147$ and $0.502$), comparable precision in parameter estimates (mean empirical SE of $0.012$ and $0.014$ for the MHMM and the HMM, respectively), and it exhibited a better coverage of the true group-level transitions (mean 95\% coverage of $.712$ and $.126$ for the MHMM and the HMM, respectively). This trend persisted in Scenario 4, where the MHMM displayed a mean absolute bias of $0.027$ (compared to $0.027$ in the HMM), mean absolute relative bias of $0.187$ (compared to $0.476$ for the HMM), mean empirical SE of $0.021$ (compared to $0.020$ for the HMM), and mean 95\% coverage of $.750$ (compared to $.117$ for the HMM) over the transition probabilities. Similar to the findings in Scenario 2, the 95\% credible intervals in the HMM remained excessively narrow, indicating overconfidence in the model estimation, even after bias correction, leading to persistent under-coverage.

\textit{Variance components}.

In Scenarios 1 and 3 (Tables \ref{table:supp_1} and \ref{table:supp_3}), where no between-heterogeneity is present in the transition distribution, the MHMM consistently overestimated the --absence of-- heterogeneity. This resulted in a mean upward bias of $0.363$ (SD $=0.159$) and $0.956$ (SD $=0.436$) in the estimation of transition variance components (the variance of the individual-specific state-specific random intercepts) for each scenario, coupled with substantial under-coverage of their 95\% credible intervals (as the value zero was never covered). Notably, the substantial bias in the between-individual variability of intercepts in the logit domain has a negligible impact on the between-individual heterogeneity on the probability scale, i.e., the variance between the transition probabilities calculated \textit{ad hoc}, which remains relatively accurate. Specifically, the upward bias is considerably smaller for the \textit{ad hoc} variance between individuals' transition probabilities, with means of $<0.001$ (SD $=0.001$) and $0.002$ (SD $=0.001$) for Scenarios 1 and 3, respectively, although the value zero remains excluded from the 95 \% credible intervals, resulting in null coverage. However, this poses a limitation on the measurement of the between-individual variability itself.

When between-heterogeneity is present only in the transition distribution (Scenario 2), the estimation of transitions' variance components improves in the MHMM, although an upward bias is still present. The transitions' variance components in the logit scale are estimated with a mean upward bias of $0.085$ (SD $=0.043$) and a mean coverage of the 95\% credible intervals of $.973$ (range: $.952$-$.992$). However, Scenario 4, with heterogeneity in both components of the model, exhibited the worst estimation performance of the transitions' variance components with a mean upward bias of $1.025$ (SD $=0.570$) and a mean coverage of the 95\% credible intervals of $.451$ (range: $.063$-$.933$). Similarly as before, the seemingly large biases in the logit scale translate to relatively small deviations in the estimation of between-heterogeneity calculated \textit{ad hoc} as the variance in the individual transition probabilities, with mean absolute biases of $0.002$ and $0.003$ for Scenarios 2 and 4, respectively, and mean coverage of the 95\% credible intervals of $.829$ (range: $.684$-$.992$) and $.766$ (range: $.242$-$.992$).

\subsubsection{Emissions}

\begin{figure*}[ht]
\centering
\includegraphics[width=0.7\linewidth]{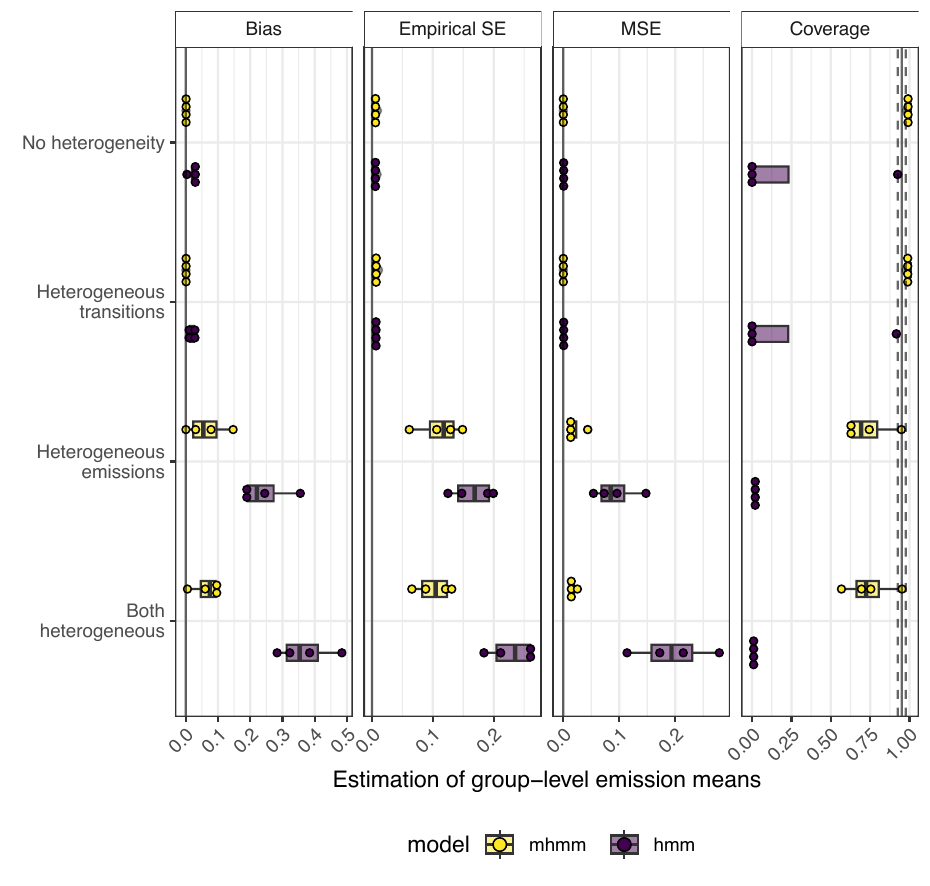}
\caption{Estimation performance group-level emission means $\bar{b}_{ij}$ on the MHMM (in yellow) and the HMM (in purple) over the four conditions of between-individual heterogeneity (no heterogeneity, heterogeneous transitions, heterogeneous emissions, both heterogeneous). Dots represent \textit{maximum a posteriori} parameter point-estimates, with overlaying box-plots to facilitate interpretation.}
\label{fig:emiss_fixed}
\end{figure*}

\textit{Group-level means}. Figure \ref{fig:emiss_fixed} summarizes the mean bias, empirical standard error (SE), and coverage of the Monte Carlo simulation study on the estimation of emission group-level parameters for the MHMM and the HMM (for exact values in all parameters, see Tables \ref{table:supp_1}--\ref{table:supp_4}). The results in Figure \ref{fig:emiss_fixed} generally indicate a good estimation performance of the group-level Poisson parameters by the MHMM, outperforming the HMM in every scenario.

In Scenarios 1 and 2, where no between-heterogeneity is present in the emissions, no substantial biases were observed in estimating group-level emission means, with absolute bias of $\leq0.001$ for the MHMM (compared to $\leq0.035$ for the HMM). Both models presented precise estimations, with an empirical standard error under $0.011$ and an MSE lower or equal than $0.001$. The mean coverage of the 95\% credible intervals was substantially better in the MHMM ($.995$, range: $.976$-$1.000$) compared to the HMM ($.230$, range: $.000$-$.925$), although over-coverage is present in the former.

In Scenarios 3 and 4, where between-heterogeneity is present in the emissions, both models show larger biases compared to the previous scenarios, although the MHMM consistently showed better performance than the HMM. In Scenario 3, the MHMM presented a lower mean absolute bias (MHMM: $0.064$, HMM $0.244$), higher precision in parameter estimates (mean empirical standard error of $0.111$ and $0.165$ for the MHMM and the HMM, respectively), lower average MSE (MHMM $0.021$, HMM $0.093$), and it exhibited a better coverage of the true group-level Poisson emission means (mean 95\% coverage of $.737$ (range: $.616$-$.948$) and $.019$ (range: $.004$-$.036$) for the MHMM and the HMM, respectively). Similar trends occurred in Scenario 4, where the MHMM displayed a mean absolute bias of $0.064$ (compared to $0.368$ in the HMM), mean empirical standard error of $0.101$ (compared to $0.229$ for the HMM), lower average MSE (MHMM $0.016$, HMM $0.195$), and mean 95\% coverage of $.742$ (range: $.567$-$.952$), compared to $.010$ (range: $.000$-$.020$) for the HMM, over the Poisson emission means.

\textit{Variance components}.

In Scenario 1 and Scenario 2 (Tables \ref{table:supp_1} and \ref{table:supp_2}), the absolute bias for all emission variance components was small ($\leq0.004$), indicating that the MHMM was able to deal adequately with the lack of heterogeneity in the emission means. Parameter estimation was precise (empirical standard error $<0.001$ for all parameters), and the MSE extremely low ($<0.001$). Since the distributional assumptions of the Bayesian estimation used an Inverse Gamma distribution to sample $\tau_{ki}$, which has a low density around zero, even the small positive bias $\leq 0.004$ led to a complete under-coverage when no between heterogeneity is present in the emission distribution ($ \tau_{ki} = 0.000$).

In Scenario 3 and Scenario 4 (Tables \ref{table:supp_3} and \ref{table:supp_4}), the MHMM exhibits a moderate negative bias for the emission variance components (ranging $-0.330$ to $-0.069$), which indicates that the model underestimates the between heterogeneity in the individual emission means. Parameter estimation was lower than in Scenarios 1 and 2 (empirical standard error range: $0.028$-$0.160$), and the MSE higher (range: $0.005$-$0.116$). The MHMM presents varied levels of coverage over the states in these two scenarios, depending on the size of the bias of each parameter (range: $.000$-$.960$). However, all values for the bias-corrected coverage ranging $.896$ to $.964$, suggesting an adequate level of confidence of the credible intervals.

\subsubsection{Decoding}

To conclude, we examined the state decoding accuracy of a multilevel hidden Markov model (MHMM) and a hidden Markov model (HMM) under the four scenarios. The mean evaluation metrics for each scenario are presented in Table \ref{table:decoding}, along with their 95\% confidence intervals.

\begin{table*}[ht]
\small
    \caption{Mean evaluation metrics for the results of the Monte Carlo simulation study on the state decoding accuracy.}
    \centering
    {
    \begin{tabular}{ccccccc}
    \toprule
        Scenario & Metric & \multicolumn{2}{c}{MHMM} & ~ & \multicolumn{2}{c}{HMM}   \\
        \cline{3-4}\cline{6-7}
        ~ & ~ & Mean & 95\%CIs & ~ & Mean & 95\%CIs  \\ \hline
        \multirow{3}{*}{\makecell{No\\heterogeneity}} & Bal. Acc & .997 & .996-.997 & ~ & .997 & .996-.997  \\
        ~ & F1 score & .996 & .995-.996 & ~ & .996 & .995-.996  \\
        ~ & Cohen's $\kappa$ & .991 & .989-.992 & ~ & .991 & .989-.992  \\ \hline
        \multirow{3}{*}{\makecell{Heterogeneous\\transitions}} & Bal. Acc & .997 & .996-.997 & ~ & .997 & .996-.997  \\ 
        ~ & F1 score & .995 & .994-.996 & ~ & .994 & .993-.996  \\
        ~ & Cohen's $\kappa$ & .992 & .990-.993 & ~ & .991 & .988-.992  \\ \hline
       \multirow{3}{*}{\makecell{Heterogeneous\\emissions}} & Bal. Acc & .879 & .845-.911 & ~ & .793 & .704-.843  \\ 
        ~ & F1 score & .822 & .768-.871 & ~ & .701 & .578-.770  \\
        ~ & Cohen's $\kappa$ & .765 & .696-.830 & ~ & .625 & .474-.706  \\ \hline
        \multirow{3}{*}{\makecell{Heterogeneous\\transitions\\\& emissions}} & Bal. Acc & .875 & .838-.907 & ~ & .773 & .678-.838  \\ 
        ~ & F1 score & .820 & .759-.872 & ~ & .669 & .547-.758  \\ 
        ~ & Cohen's $\kappa$ & .751 & .668-.818 & ~ & .573 & .403-.692 \\
        \bottomrule
    \end{tabular}
    }
    \label{table:decoding}
\end{table*}

For Scenario 1 and Scenario 2, both the MHMM and HMM models achieved excellent levels of state decoding accuracy, with mean balanced accuracy Balanced Accuracy, F1 score, and Cohen's $\kappa$ values ranging from $.991$ to $.997$ for the MHMM, and $.990$ to $.997$ for the HMM. No substantive differences were detected between the decoding metrics of the two models. The 95\% credible intervals were narrow, indicating that the results were precise.

In contrast, Scenario 3 and Scenario 4 (with between heterogeneity in the emission distribution) were more challenging to estimate, resulting in lower state decoding accuracy for both models. Notably, the MHMM outperformed the HMM in both scenarios, achieving statistically significant improvements over the Balanced accuracy, the F1 score, and Cohen's $\kappa$ of $10.6\%$, $15.3\%$, and $21.31\%$ in Scenario 3, and $13.5\%$, $21.35\%$, and $31.6\%$ in Scenario 4, respectively. Although the inclusion of between-individual heterogeneity in the transition distribution alone did not show an impact on the decoding accuracy (Scenario 2), the data generation process with between heterogeneity in both components of the model (Scenario 4) led to a lower decoding accuracy over the three metrics in the HMM when compared to the scenario only including between heterogeneity in the emission distribution (Scenario 3). The same pattern was not detected for the MHMM, whose decoding accuracy was equivalent in Scenarios 3 and 4. Finally, we notice that the confidence intervals for the MHMM and HMM models in Scenarios 3 and 4 were wider than in Scenarios 1 and 2, which indicates greater variability in the decoding accuracy over the simulation repetitions.

\section{Empirical applications}
\label{sec:emp_applications}

In this section, we demonstrate the utilization of the multilevel Hidden Markov Model (HMM) to investigate the underlying dynamics that govern complex multivariate count data in two distinct practical scenarios. The first example involves the identification of different types of diving behaviors exhibited by pods (i.e., groups) of long-finned pilot whale in the natural habitat, exploring variations in their latent behavioral dynamics. By incorporating individual random effects into the model, it becomes feasible to evaluate the unique differences among the pods of whales. The second practical application serves as an illustration of how to identify neural states from multi-electrode recordings of motor neural cortex activity in a macaque monkey within an experimental setup, where the observational units are the individual trials. In this case, individual random effects within the model act as a tool to reliably unveil neural states despite the heterogeneity between trials, ensuring that the meaning of the states remains consistent across them.

\subsection{Pilot whale data}
\label{sec:whales}

We present an empirical demonstration of the multilevel Hidden Markov Model (HMM) discussed in section 2. This demonstration employs behavioral data obtained from 15 long-finned pilot whales in the Vestfjord region of Norway, originally analyzed in the work of \citet{isojunno_individual_2017} using a mixture HMM with discrete random effects. Comprehensive details can be found in the original study. The dataset comprises several distinct data streams (e.g., dive depth, ground speed, head pitch, etc.) associated with "Foraging", "Exploratory", "Crowded", and "Directed" diving behaviors exhibited by these 15 individuals. The number of recorded occasions varies between 50 and 254, with a median of 148, as reported by \citet{mcclintock_worth_2021}. \citet{mcclintock_worth_2021} demonstrated that the inclusion of continuously distributed individual random effects contributes to the stability of results in cases where convergence is challenging. This approach also aids in identifying variations in transition distributions among the whales. Although the complete dataset includes multiple observational variables characterizing dive characteristics, our focus narrows down to three variables that align more closely with the definition of a count variable due to their skewed distribution. These variables are the \textit{duration of dives} in minutes (rounded to the nearest integer), the \textit{depth of dives} in meters (rounded to whole units), and the \textit{size of the whale group} (pod) during each dive. It is important to note that we employ this empirical dataset primarily as a means to showcase the insights that can be derived using our novel model. However, we do not assert that a Poisson distribution is the best theoretical fit for these variables. For the sake of result interpretation, we have fixed the number of states at four, a choice consistent with the model selection findings in \citet{isojunno_individual_2017} and confirmed by \citet{mcclintock_worth_2021}.

Same as for the simulation study, we fit the multilevel HMM with a Poisson-lognormal distribution described above (for details, see Section \ref{sec:model_fitting}). The model was fit with 4000 iterations, with the first 2000 being discarded to dissipate the effect of starting conditions (burn-in). Contrary to the simulation study, here we specified weakly informative hyper-priors centred on on the group-level means found in \citet{isojunno_individual_2017} to ensure the stability of the Monte Carlo Markov chains. As starting values for the parameter chains, we used random values centred on the same group-level values used for the weakly informative hyper-priors. The sequence of most likely states given the behavioural data, along with each state’s forward probabilities were determined for each trial with the Viterbi algorithm \citep{Forney1973, Viterbi1967} based on whale-specific parameters. Finally, we relied on posterior predictive checks (PPCs) to assess the adequacy of the model fit to the empirical data. For code to reproduce the analyses refer to \url{https://doi.org/10.5281/zenodo.10834837}.

\begin{figure}[ht]
\centering
\includegraphics[width=\linewidth]{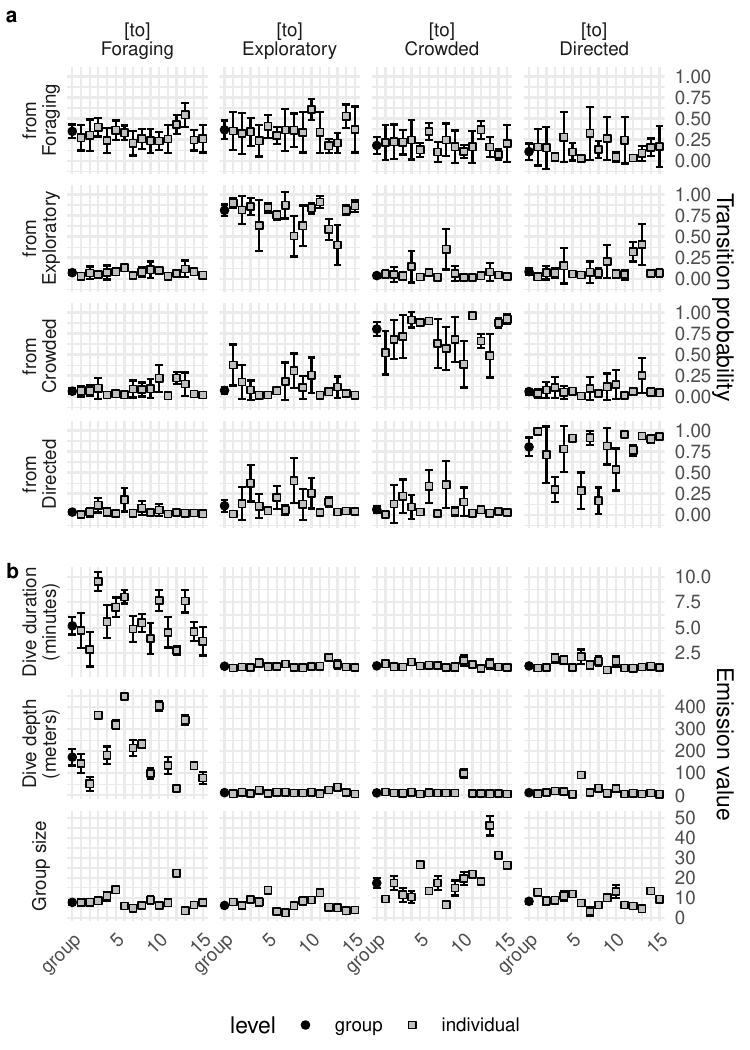}
\caption{Group-level (in black) and whale-specific (in grey) transition (a) and emission (b) \textit{maximum a posteriori} parameter point estimates, displayed with standard error. Whales exhibit heterogeneous levels of variability across states and dependent variables. Notice that the uncertainty is larger for Foraging dives, since they occur less frequently in the data.}
\label{fig:whale_components}
\end{figure}

The multilevel HMM produced stable results for the model with four states, which generally reproduced the group-level patterns previously found by \citet{isojunno_individual_2017}. The composition of the four states in Figure \ref{fig:whale_components}b go in line with the classification on "Foraging", "Exploratory", "Crowded", and "Directed" dive behavior in pilot whales. Foraging dives diverge clearly from the other three type of dives, and they are characterized by a relatively long duration (median = 5.15 minutes), high depths (median=167 meters) and on par group size (median = 7.73 individuals) at the sample-level. Meanwhile, the differences between Exploratory, Crowded, and Directed types of dives is more subtle: the three types of dives share a shorter median duration (1.20, 1.26, and 1.19 minutes, respectively), reach lower median depths (11.97, 11.10, 11.52 meters, respectively), although their median group sizes appear to differ more (6.24, 17.15, and 8.36 individuals per pod). Global state decoding with the Viterbi algorithm revealed that Exploratory dives were the most frequent dive type (35.8\% of all dives), followed by Directed (34.7\%) and Crowded dives (24.8\%), with Foraging dives consisting of only 4.6\% of the data.

\begin{figure*}[ht]
\centering
\includegraphics[width=0.7\linewidth]{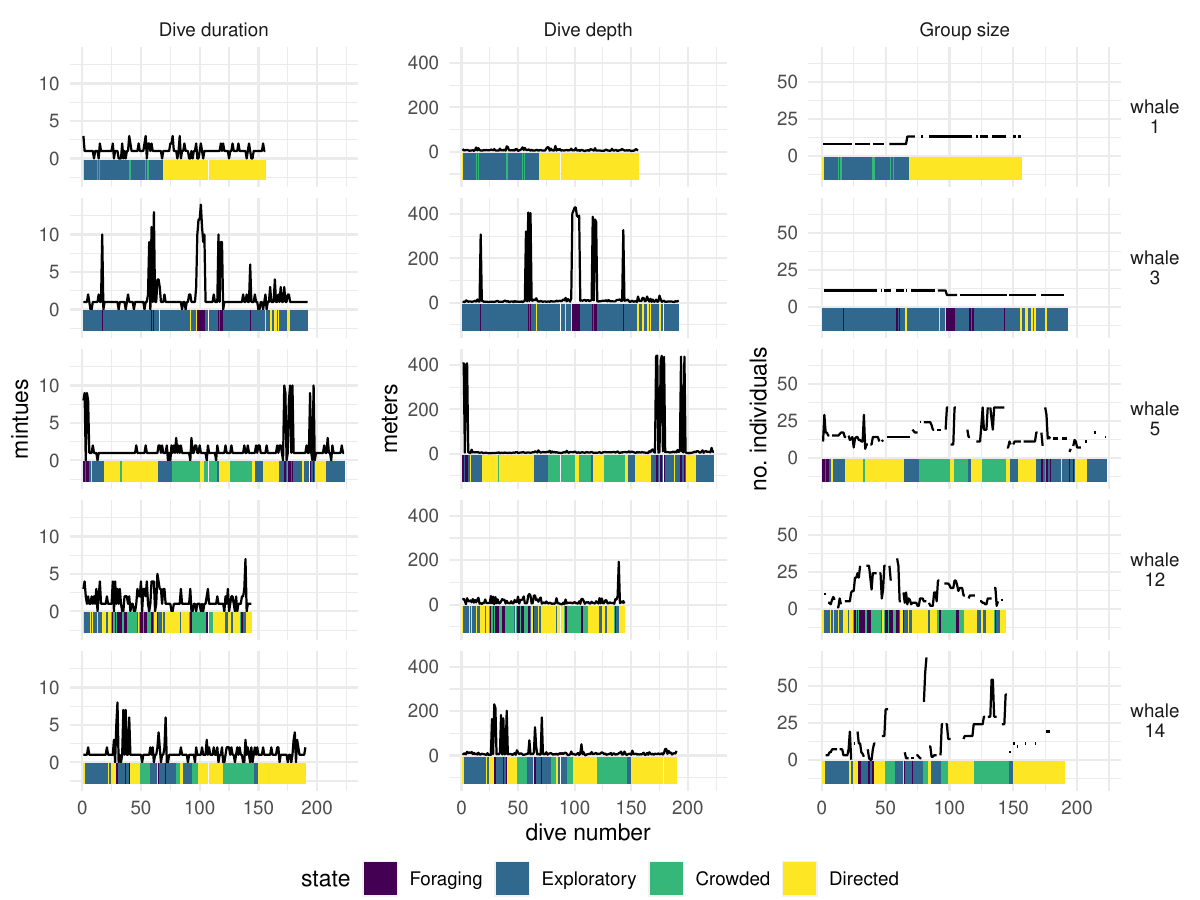}
\caption{Decoded actograms with the mostly likely behavioural state for each dive number shown for five exemplary whales (in rows) and the three observed time series (Dive duration, Dive depth, Group size; in columns).}
\label{fig:whale_decoding}
\end{figure*}

The average switching dynamics between dive types (hidden states in this model) is shown in Figure \ref{fig:whale_components}a. Foraging dives are generally more likely to be followed by Exploratory or Crowded dives, than by Directed dives. They were also the only type of dives in which switching to a different type of dive was more likely than exhibiting a new display of Foraging dive. Meanwhile, Exploratory, Crowded, and Directed dives tended to have a higher temporal persistence (i.e., larger self-transition probabilities), and were more likely to be followed by a same type of dive behavior. The inclusion of individual random effects revealed evidence of between-individual differences in both components of the HMM, as depicted in Figure \ref{fig:whale_components}a. Whether these differences can be attributed to environmental, or individual differences on the "personality" of animals is not further explored in this application. The recordings of five of the whales on Figure \ref{fig:whale_decoding} serve as example of individual-specific deviations on the emission distribution: whales are assigned to different dive states due to divergences on the expected values for one or other type of dive (i.e., different baselines over the states for a dependent variable). Notwithstanding, the general trends on the relation between the Poisson parameters over the states in each dependent variable are relatively stable across whales, as visualized by mostly parallel lines in Figure \ref{fig:whale_emiss_linked}.

\begin{figure*}[ht]
\centering
\includegraphics[width=0.7\linewidth]{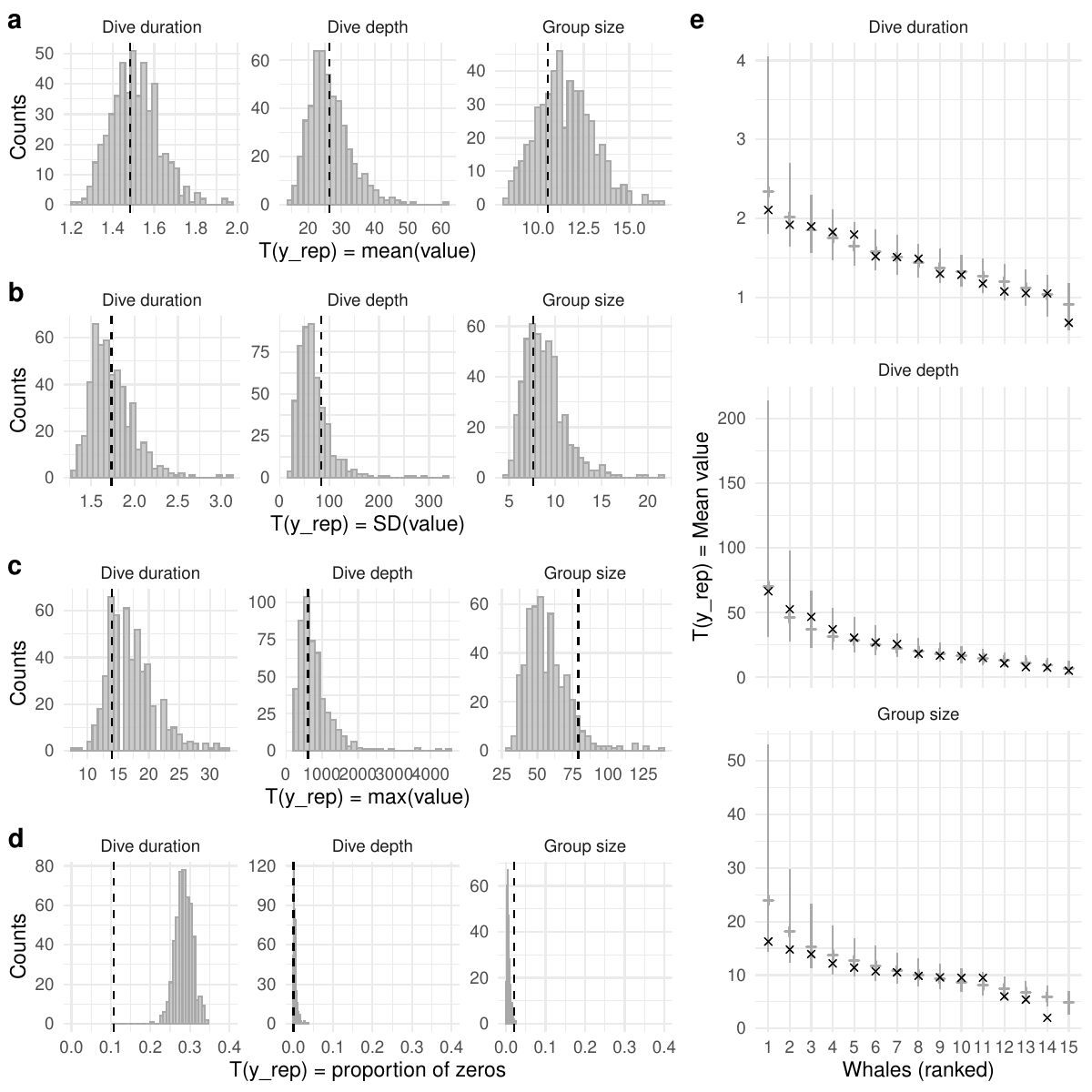}
\caption{The data generated by the MHMM-PLN generally recover the patterns observed on the empirical behavioural data, which indicates good fit. (a--d) Histograms show the distribution of four group-level summary statistics $T^{rep}_{(y)}$ calculated for the Dive duration, Dive depth, and Group size over 500 simulated data sets: (a) mean, (b) standard deviation, (c) maximum, and (d) proportion of zeros. Dashed lines indicate the scores of the summary statistics $T^{emp}_{(y)}$ of the empirical whale data. (e) Displays the ranked distribution of whale-specific emission means for the 15 whales in the simulated repetitions, along with crosses indicating the whale-specific means in the empirical behavioural data.}
\label{fig:whale_ppcs}
\end{figure*}

To conclude, we demonstrate the process of assessing model fit through posterior predictive checks (PPCs). The idea is to check whether the patterns and characteristics seen in the observed data are also present in the simulated data. This is done simulating a large number of new synthetic datasets from the fitted model, to obtain a posterior predictive distribution for each simulated whale. Then, the simulated data is then compared to the observed data at the individual or group level for a number of statistics (e.g., mean counts, proportion of zero counts, etc). Figures \ref{fig:whale_ppcs}a--d show the PPCs for the distribution of four statistics taken over the aggregated data for the three dependent variables included in the model: these are the mean, standard deviation, and maximum values observed over the 500 synthetic data sets, and the proportion of zeros found in them. Overall, the posterior predictive distribution of the statistics generally suggest a good model fit at the aggregated level. However, the model tends to overestimate the proportion of dives with a duration zero (Figure \ref{fig:whale_ppcs}d), and, to a lesser extent, it underestimates the proportion of dives with depths and group sizes of value zero (Figure \ref{fig:whale_ppcs}d). The distributions of individual specific means on the synthetics data sets in Figure \ref{fig:whale_ppcs}e do not show substantial deviations from goodness of fit.

\subsection{Neural activation data}
\label{sec:monkeys}

In this section we present a second empirical application of the model to illustrate how it can be used to uncover neural states from \textit{in vivo} electrophysiological experiments while accounting for individual differences between multiple trials. Without going into too much detail, one  of the current views in neurophysiology supports the hypothesis that neural populations, rather than single neurons, may be the fundamental unit of cortical computation \citep{saxena_towards_2019}. Although basic HMMs have been successfully applied to these types of data to uncover neural states, analysing chronically recorded neural population activity can be challenging not only because of the high dimensionality of activity but also because of changes in the signal that may occur due to neural plasticity \citep[e.g., task-induced plasticity,][]{dayan_neuroplasticity_2011} or not \citep[e.g., electrode drift or degradation,][]{barrese_scanning_2016,welle_ultra-small_2020}. From a statistical modelling perspective, these changes in the recording signal cause variation over trials, which is ignored when using conventional HMM. When applying the multilevel HMM, this variation can be accommodated in the model. 

\begin{figure*}[ht]
\centering
\includegraphics[width=0.7\linewidth]{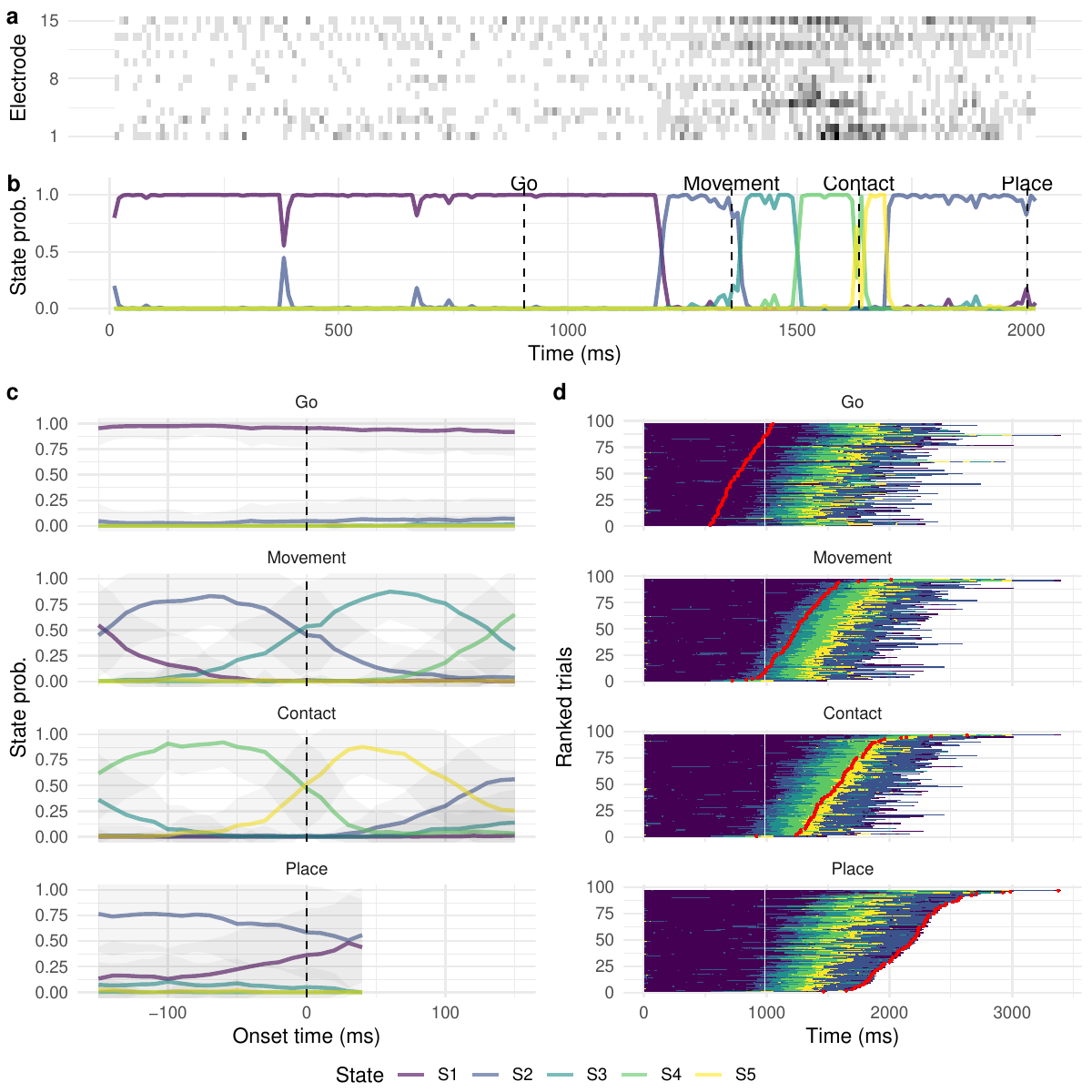}
\caption{MHMM uncovered states align with phases of the reaching task. (a) Raster plot of a single example trial with the spiking activity over the 15 electrodes. (b) Forward probabilities for each state given for the trial in Figure \ref{fig:monkey_results}a. (c) Mean forward probabilities (solid lines, shaded areas represent standard error) for each state averaged over trials and aligned to each behavioural event onsets (go: go signal, movement: reaching movement, contact: object grasp, place: object placing). (d) Stacked state decoding over trials sorted by behavioural event times. Each row shows the most likely state for each trial for every time step with the corresponding event time overlaying in red.}
\label{fig:monkey_results}
\end{figure*}

To show how to extract neural states while accounting for different sources of variability, we use a selection of the data previously analysed in \citet{kirchherr_bayesian_2022}, consisting of the recordings of neural activity of two monkeys trained to reach, grasp, and place an object after receiving a sound stimulus in an experimental set up, while they perform such task. Here, the dependent variables are the number of neural activations (spike counts) per 10 millisecond bin produced while the monkeys perform the reaching task, recorded with multi-electrode probe surgically inserted in its primary motor cortex (M1) (see Figure \ref{fig:monkey_results}a for an example of a trial). Although \citet{kirchherr_bayesian_2022} considered over 900 trials combined between two monkeys, for the illustrative purpose of this application we focus on the spike counts recorded with 15 electrodes in 97 experimental trials performed in the span of two days on a single macaque monkey (\textit{Macaca mulatta}) (Day 1--2, monkey 2, electrodes 1--15 in the original data; we refer the reader to \citet{kirchherr_bayesian_2022} for further information on the collection of the data set). Here, each trial will be used as an independent individual in the model, with the spike counts over time recorded by the 15 electrodes, the dependent variables in the model, and the number of hidden states fixed to 5 (the same as in the original study). Notice that whereas the previous empirical application was used to explore the individual differences in the behaviour of a relatively small number of individuals (whales) observed in their natural environment, here, we focus on uncovering the forward probabilities of hidden states on different trials with respect to a controlled experimental stimulus.

We followed the same strategy as in the first empirical example in Section \ref{sec:whales} to train a MHMM Poisson-lognormal with 5 hidden states, using the group-level parameter point estimation in \citet{kirchherr_bayesian_2022} for the starting values and weakly informative hyper-priors. We extracted the global decoding along with the state forward probabilities obtained with the Viterbi algorithm \citep{Forney1973, Viterbi1967}, and we relied on posterior predictive checks (PPCs) to assess the adequacy of the model fit to the empirical data. The event onset times of four consecutive behavioral clues (go signal, arm movement, object contact, and object placing) were used to rank the experimental trials and provide an interpretation to the hidden states uncovered. For code to reproduce the analyses refer to \url{https://doi.org/10.5281/zenodo.10834837}.

The model with five hidden states was able to provide stable results and reproduce the general pattern of results obtained for a much larger number of trials (383) over a larger number of days (10) using a larger number of electrodes (25) in the original study.

The five hidden states uncovered with the MHMM generally show a good temporal alignment with different stages of the reaching task performed by the monkey, as exemplified in Figure \ref{fig:monkey_results} showing the electric activity of a trial (Figure \ref{fig:monkey_results}a, raster plot), the estimated forward probabilities for the same trial (Figure \ref{fig:monkey_results}b), and the average forward probabilities over all trials aligned using the behavioral onsets (Figure \ref{fig:monkey_results}c). Notice that, hidden states in this context represent groups of neurons whose electrical activity changes synchronically during a period of time, conforming what is known as a transient neural state, which may be involved in the orchestration of a motor task. Here, the first neural state (State 1) appears to represent a basal state of general uniform activity in the neurons recorded with the 15 electrodes in the example (Figure \ref{fig:monkey_results}c, Go panel; see also Figure \ref{fig:monkey_group_emiss} for the group-level emission distribution). The forward probability of the second neural state (State 2) becomes the most likely at about 200 ms prior to the hand movement onset (at which the monkey starts to reach for the object) and peaks at 150 ms, after which it recedes, giving place to the third neural state (State 3) (Figure \ref{fig:monkey_results}c, Movement panel). The onset of the hand movement coincides with the switching between States 2 and 3, with the later peaking at about 75 ms after the movement onset (Figure \ref{fig:monkey_results}c, Movement panel). The fourth neural state (State 4) precedes the object contact, peaking approximately 75 ms before it (Figure \ref{fig:monkey_results}c, Contact panel). The onset of the object contact marks the switching between States 4 and 5, with the later exhibiting a shorter duration (with a peak at about 50 ms after the object contact onset; Figure \ref{fig:monkey_results}, Contact panel). Approximately 125 ms after the object contact onset, the second neural state (State 2) becomes the again the most likely state, and it remains that way until right after the placement of the object (Figure \ref{fig:monkey_results}, Contact and Place panels). Finally, the temporal neighborhood of the object placing appears to be accompanied by an increase in the likelihood of the first neural state (State 1) in at least some of the trials (Figure \ref{fig:monkey_results}c, Place panel), recapitulating a basal activity over the fifteen electrodes for the remainder of the them (e.g., see Figure \ref{fig:monkey_results}a). The same temporal alignment between the average forward probabilities across and the behavioral landmarks was observed in most of the trials, as depicted in the global state decoding for the 97 trials in the data ordered by the corresponding landmark onset in Figure \ref{fig:monkey_results}d.

\begin{figure*}[ht]
\centering
\includegraphics[width=0.5\linewidth]{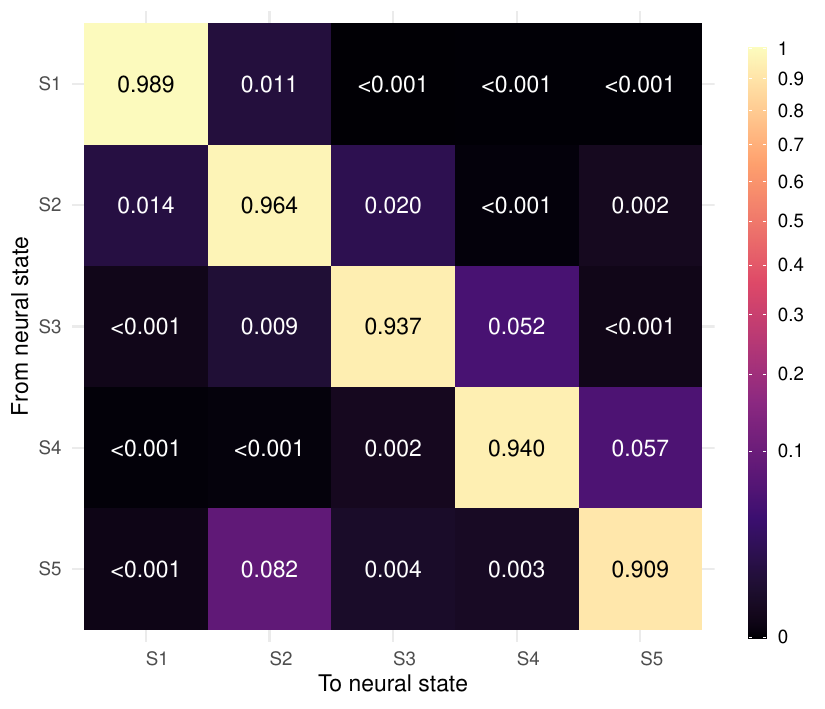}
\caption{Group-level transition probability matrix (\textit{maximum a posteriori} estimates) describing the dynamics of switching between states. The pattern of the off-diagonal transition probabilities suggests a sequential left-to-right transition path, with the next state always more likely than the previous state(s), except for S5.}
\label{fig:monkey_group_gamma}
\end{figure*}

The switching dynamics between states described above are also captured by the group-level transition probability matrix in Figure \ref{fig:monkey_group_gamma}. The off-diagonal transitions reflect the consecutive nature of the states, with the highest off-diagonal probability of moving from one state to the immediate consecutive (e.g., $S1 \rightarrow S2$, $S2 \rightarrow S3$, and so on) with the notable exception of state 5, which is more likely to switch to state 2. Notice that, although such transition probability matrix suggests a left-to-right process, backward transitions between states are still allowed (and in fact are more likely than transitions between other states). The high self-transition probabilities obtained for the five states go in line with states that are persistent over time. Global decoding with the Viterbi algorithm reveals that state 1 occurs most frequently in the data (51.4\%), followed by state 2 (25.3\%), state 3 (8.6\%), state 4 (8.4\%), and state 5 (6.3\%). The mean empirical duration of the states varied substantially over the decoded trials,  placing state 1 as the most persistent ($398 \pm 454$ ms), followed by state 2 ($116 \pm 134$ ms), state 4 ($109 \pm 94$ ms), state 3 ($87 \pm 71$ ms), with state 5 ($62 \pm 52$ ms) the shortest lived following the object contact. A substantial between-trial variability was also observed for their emission parameters which warrants the adoption of a multilevel approach, as visualized in Figure \ref{fig:monkey_trial_emiss}.

\begin{figure*}[ht]
\centering
\includegraphics[width=0.7\linewidth]{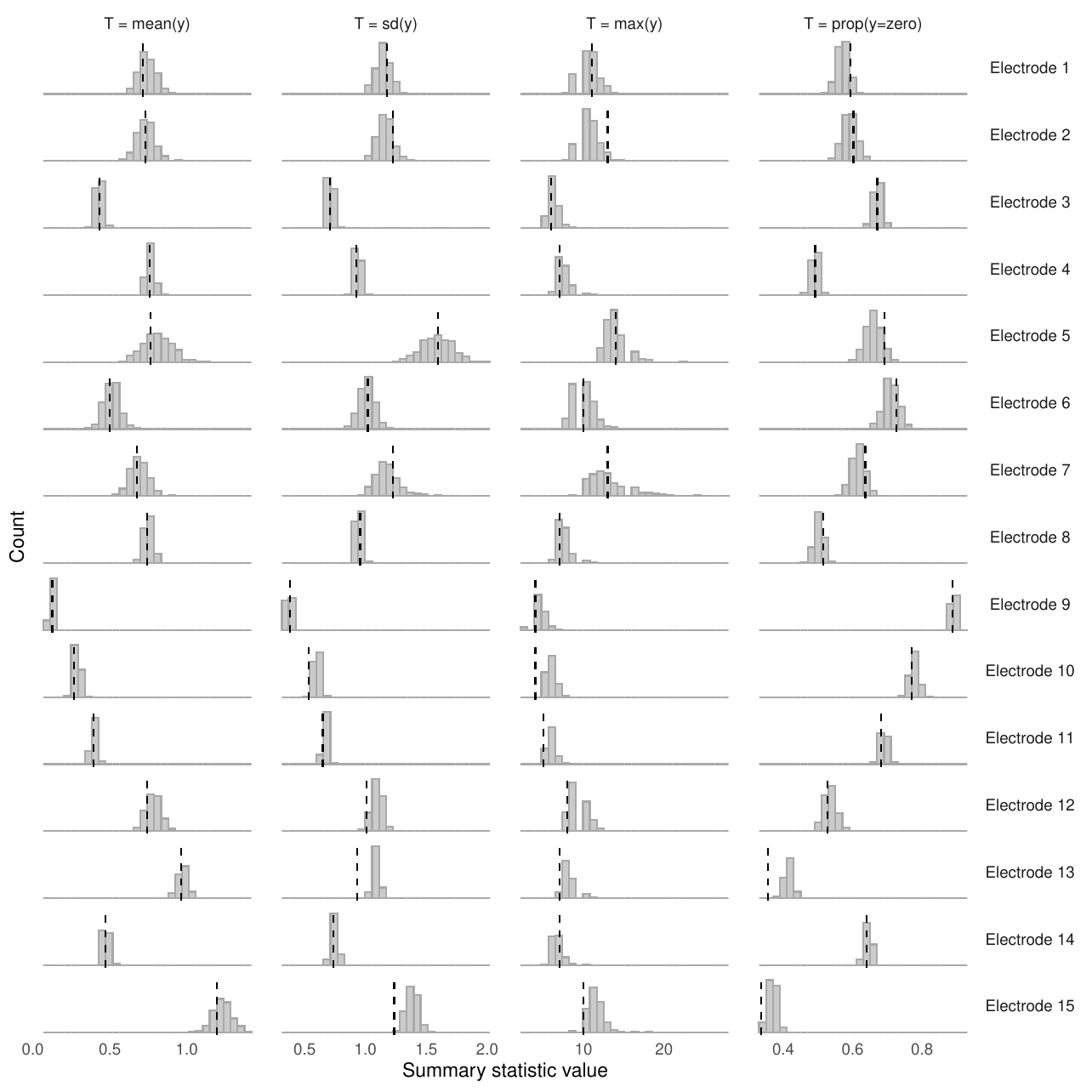}
\caption{Data generated by the MHMM-PLN recovers the patterns assessed on the aggregated empirical data, which indicates good fit. Histograms show the distribution of four summary statistics $T^{rep}_{(y)}$ (mean, standard deviation, maximum, and proportion of zeros) of the synthetic spike counts for the 15 electrodes over 500 simulated data sets. Dashed lines indicate the scores of the summary statistics $T^{emp}_{(y)}$ of the empirical data.}
\label{fig:monkey_ppc}
\end{figure*}

The posterior predictive checks in Figure \ref{fig:monkey_ppc} did not show substantial deviations from goodness of fit, as the model was able to reproduce the values of the summary statistics taken on the aggregated empirical data for most of the electrodes. Extreme posterior predictive values were only obtained for electrode 2 (underestimation of maximum spike counts), electrode 10 (overestimation of the variability and the maximum spike counts), and electrode 13 (overestimation of variability and proportion of zeros in the spike counts). Notwithstanding, considering the 15 dependent variables and the fact that previous research has shown a strong beneficial effect for including multivariate data on the estimation performance of the model \citep[e.g.,][]{mildiner_moraga_go_2023, wurpts_is_2014}, we expect the state decoding to be robust to these type of deviations.

\section{Discussion}
\label{sec:discussion}

In this study we developed and made available to the community a Bayesian multilevel hidden Markov model with a Poisson-lognormal emission distribution to analyse (multivariate) count time-series of multiple subjects or clusters at the time. We showed that it is possible to include continually distributed individual random effects in both components of the model, and described how to perform the estimation of the parameters in the Bayesian framework. By means of a small Monte Carlo simulation study, we also showed that the multilevel HMM performs equally well than an equivalent (single-level) HMM without random-effects over data generation conditions where no between-individual variability is present in the data generation mechanism, and outperforms it when individual heterogeneity is present. Finally, we demonstrated how to extract insights from two real data using the multilevel HMM with two empirical applications. Overall, these results suggest that the multilevel HMM is a safer choice than the single-level HMM if one is agnostic about the presence of between individual heterogeneity in the data.

A key finding from this study underscores the importance of considering individual heterogeneity when modeling emission distribution parameters. Individuals and groups exhibit variations in how their underlying states manifest (i.e., externalize), influenced by factors like age, sex, or other unobservable individual-specific characteristics. The results presented here reveal that neglecting this heterogeneity—incorrectly assuming a uniform set of parameters for all individuals ("complete pooling")—not only leads to biased estimates of emission parameters (i.e., lower or larger expected counts given the state) but also results in significantly poorer accuracy in decoding the underlying process (i.e., the sequence of hidden states) as demonstrated by a worse performance with the single-level HMM when compared to the multilevel HMM. While previous empirical studies have supported addressing individual variation in emission distributions \citep[e.g., ][]{Altman2007, langrock_flexible_2012, Rueda2013, kirchherr_bayesian_2022}, our study provides the first evidence of the profound impact of neglecting inter-individual heterogeneity on overall estimation accuracy. Our findings also align with the work of \citet{mcclintock_worth_2021}, which previously demonstrated that neglecting individual variation in the latent process (specifically, in transition probabilities) has only a minimal impact on model accuracy and the overall estimation of group-level emission means. The current results echo those of \citet{mcclintock_worth_2021}, as failure to account for heterogeneity in the transition distribution with the single-level HMM resulted in nearly perfect decoding accuracy, comparable to the multilevel HMM. Consequently, the study emphasizes that incorporating heterogeneity in emission distribution parameters has a more substantial influence on the model's decoding accuracy.

As illustrated in the first empirical example in Section \ref{sec:whales}, one of the primary benefits of using the MHMM is its capability to assign distinct parameters to each unit in the datasets. This proves especially advantageous for researchers who aim to investigate individual variations, focus on personality traits, individual dynamics, and related aspects. While some of these differences can be partially explored using individual covariates, certain traits contributing to inter-individual disparities may remain unobservable directly or lack a suitable means of measurement. In other instances, the incorporation of individual-specific random effects may merely serve a practical purpose. For example, in the second empirical application in Section \ref{sec:monkeys}, the inclusion of random effects was done with the purpose of enhancing the model's decoding accuracy and making estimation possible over a complex data set by controlling for the unexplained heterogeneity between trials due to an experimental set up. This doesn't necessarily imply a special interest in studying the individual properties of the trials, but rather underscores the importance of considering this nuance to prevent any detrimental effects on the results of the analyses.

The results of the main simulation also revealed that the estimation of the individual specific random effects of the transition distribution appear to be especially hard, since not all parameters achieved convergence in every repetition, and the variance components $\sigma_{ij}$ tended to be overestimated by the model. This result goes in line with the previous evidence for multilevel models and multilevel time-series \citep[e.g.,][]{asparouhov_dynamic_2018, hox_multilevel_2018, Landau2013, mcclintock_worth_2021}. Although previous literature suggest that 50 individuals may be enough for accurate estimation of variance components, such recommendation is for linear regression models without a temporal component. The number of individuals required, though, appears to be larger for MHMMs \citep[][]{mcclintock_worth_2021, mildiner_moraga_go_2023}, and still remains to be explored for the MHMM-PLN. Based on the results of \citet{mildiner_moraga_go_2023} we expect that increasing the number of individuals and the number of dependent variables alleviate issues with the stability and the accuracy of on the estimation of these parameters. Interestingly, even though the MHMM-PLN tended to overestimate the level of heterogeneity between individuals (measured as variance components), it generally produced a good indication of the occurrence of heterogeneity between individuals (or the lack thereof) and their inclusion lead to an improvement over the estimation of group-level transition and emission means, and accuracy of the decoding. Furthermore, the \textit{ad hoc} empirical between variability calculated as the variance between individual-specific transition probability matrices proved to be accurate.

The overestimation of transition random effects is likely a consequence of using a scaled Inverse-Wishart distribution to sample the variance-covariance matrices of random effects, which has a low density around zero \citep{gelman_bayesian_2006, lemoine_moving_2019}. The estimation of variances can perform poorly when variances are small relative to means, constraining variances upwards and correlations downwards \citep{alvarez_bayesian_2014}. As a result, sampling from the Inverse-Wishart can lead to slight to moderate overestimation of the posterior variance-covariance matrix \citep{alvarez_bayesian_2014, lemoine_moving_2019}. In the future, a different prior distribution other than the Inverse-Wishart could be explored to sample the covariance matrix of the multivariate normal distribution $\Psi$ used to sample the individual-specific transition intercepts $\alpha_{nij}$. 

In this study, our primary objective was to provide the initial formal description of the multilevel Hidden Markov Model (HMM) for count time-series and evaluate its performance in comparison to a single-level HMM. While our model serves as a foundation, there are opportunities for practical enhancements by relaxing certain assumptions. One straightforward extension involves incorporating individual-level (i.e., time-invariant) covariates such as biological sex, body mass, or medication use \citep[e.g.,][]{Schliehe-Diecks2012, prisciandaro_identification_2019} to explain variability in individual-specific transition intercepts ($\alpha_{nij}$) and Poisson means ($b_{nki}$) or estimating the effect of an experimental treatment in one or other component of the model \citep[e.g.,][]{Altman2007, Shirley2012a, DeHaan-Rietdijk2017}. The presented multilevel HMM readily accommodates the inclusion of individual-level covariates by integrating a regression coefficient matrix into the group-level linear predictor of the multinomial logit for transitions and the linear predictor of the group-level Poisson regression for emissions, functionalities that are readily available in the accompanying R package \verb|mHMMbayes|. The model described in this study is time-homogeneous, and as such it may not capture certain processes that exhibit time-varying characteristics. Relaxing the assumption of time homogeneity could be achieved by incorporating time-varying covariates into one or both components of the model. Support for time-varying covariates in the latent process is less straightforward extension to the model, which could be achieved incorporating regression coefficients in the linear predictor of the multinomial logit at the individual-level with the transition probability matrix re-estimated for each time-step \citep[e.g.,][]{Leos-Barajas2018}. Similarly, the Poisson regression could be extended to incorporate time-varying covariates in the emission distribution. These modifications would enable a more accurate representation of latent processes that change over time, capturing phenomena like learning a new skill \citep[e.g.,][]{Visser2011} or assessing the influence of exogenous variables such as treatment or social context on the dynamics of the studied process \citep[e.g.,][]{Holsclaw2017, van_beest_classifying_2019, beumer_application_2020, photopoulou_sex-specific_2020}.

\subsection{Conclusion}
Overall, our simulation study showed that the MHMM offers advantages over the HMM for more complex data-generating processes involving heterogeneity between individuals, particularly in terms of state decoding accuracy and estimation performance of parameters of the emission distribution. Importantly, our results indicate that the MHMM should be preferred over the basic HMM when researchers do not know in advance whether they can expect between-heterogeneity in their data. To promote the adoption of our multilevel count time series analysis model within the community and facilitate its accessibility, we have incorporated the Poisson MHMM described in this work into the \verb|mHMMbayes| R package, which is publicly accessible on CRAN.

\bibliographystyle{elsarticle-harv} 
\bibliography{library}

\newpage
\appendix
\section{Supplementary materials}
\label{sec:sample:appendix}
\setcounter{figure}{0} 



Be $N$ individuals, $K$ observed time-series per individual, and $T_n$ occasions specific to individual $n$:

\begin{algorithm}[ht]
\label{mhmm_estimation_algo}
\fontsize{9}{9}\selectfont
\SetAlgoLined
\textbf{Inputs:} \\
Data. \\
$A_{0}$,$B_{0}$: initial values for the transition and emission means.  \\
$m_{0ij}$,
$K_0$,$\Psi_{0}$,$df_0$, $l_{0ki}$,
$\tau_{0ki}$,
$c_{ki}$, $d_{ki}$: hyper-parameters. \\
M: number of hidden states. \\
R: number of iterations. \\
\vspace{0.25cm}

\textbf{Initialize MCMC chains}:\\
$a_n$:=$A_0$;\\
$b_{nk}$:=$B_0$;\\
\vspace{0.15cm}
\textbf{Sample from conditional posterior distributions}:\\
\smallskip
 \For{iteration r in 1 to R}{

  \For{individual n in 1 to N}{
  
    \For{occasion t in 1 to $T_n$}{
        Calculate forward probabilities over M states and K series:
        $F_{n,t} = P(O_{nk}^{[1:t]}, S_{n}^{[t]})$;\\
    }
    \For{occasion t in $T_n$ to 1}{
        Use $F_{n}^{[1:t]}$ to backward sample: $S_{n}^{[1:t]}$;\\
    }
  }

  \For{hidden state i in 1 to M}{
    Gibbs update: $\Psi_{i.}$, $\bar{\alpha}_{i.}$;\\
    \For{time-series $k$ in $1$ to $K$}{
        Gibbs update: $\bar{b}_{ki}$,
    $\tau_{ki}$; \\
       }
    \For{individual n in 1 to N}{
    \vspace{0.1 cm}
        RW-Metropolis update: $\alpha_{ni.}$;\\
        Update: $a_{ni.}$;\\
        \For{time-series $k$ in $1$ to $K$}{
            RW-Metropolis update: $b_{nki}$; \\
        }
    }
  }
  End of MCMC iteration \textit{r}, store current parameter values
 }
 \caption{the Bayesian Poisson-lognormal multilevel HMM}

\textbf{return:} Chain samples from the posterior distributions of the parameters: $A_{n}$, $\alpha_{n}$, $\bar{\alpha}$,
$\Psi$, $B_{n}$, $\bar{b}_{k}$,
$\tau_{k}$.
\vspace{0.1cm}
\end{algorithm}


\begin{table*}

\caption{Performance metrics for the two models under conditions of no between heterogeneity}
\centering
\fontsize{10}{12}\selectfont
\begin{tabular}[t]{ccccccccc}
\toprule
\multicolumn{1}{c}{ } & \multicolumn{4}{c}{MHMM} & \multicolumn{4}{c}{HMM} \\
\cmidrule(l{3pt}r{3pt}){2-5} \cmidrule(l{3pt}r{3pt}){6-9}
parameter & bias & emp. SE & MSE & coverage & bias & emp. SE & MSE & coverage\\
\midrule
\addlinespace[0.3em]
\multicolumn{9}{l}{\textbf{Fixed effects}}\\
\hspace{1em}$\bar{a}_{11}$ & 0.003 & 0.003 & 0.000 & 0.996 & 0.000 & 0.003 & 0.000 & 0.940\\
\hspace{1em}$\bar{a}_{12}$ & -0.001 & 0.003 & 0.000 & 1.000 & 0.000 & 0.003 & 0.000 & 0.937\\
\hspace{1em}$\bar{a}_{13}$ & -0.002 & 0.001 & 0.000 & 0.937 & 0.000 & 0.001 & 0.000 & 0.940\\
\hspace{1em}$\bar{a}_{14}$ & -0.001 & 0.001 & 0.000 & 0.794 & 0.000 & 0.001 & 0.000 & 0.956\\
\hspace{1em}$\bar{a}_{21}$ & 0.001 & 0.005 & 0.000 & 1.000 & -0.001 & 0.005 & 0.000 & 0.917\\
\hspace{1em}$\bar{a}_{22}$ & 0.002 & 0.006 & 0.000 & 1.000 & 0.000 & 0.006 & 0.000 & 0.925\\
\hspace{1em}$\bar{a}_{23}$ & -0.002 & 0.004 & 0.000 & 1.000 & 0.000 & 0.004 & 0.000 & 0.956\\
\hspace{1em}$\bar{a}_{24}$ & -0.002 & 0.001 & 0.000 & 0.873 & 0.000 & 0.001 & 0.000 & 0.956\\
\hspace{1em}$\bar{a}_{31}$ & -0.002 & 0.004 & 0.000 & 0.984 & 0.000 & 0.004 & 0.000 & 0.944\\
\hspace{1em}$\bar{a}_{32}$ & 0.001 & 0.007 & 0.000 & 1.000 & -0.001 & 0.007 & 0.000 & 0.980\\
\hspace{1em}$\bar{a}_{33}$ & 0.004 & 0.008 & 0.000 & 1.000 & 0.001 & 0.007 & 0.000 & 0.964\\
\hspace{1em}$\bar{a}_{34}$ & -0.004 & 0.004 & 0.000 & 0.964 & 0.000 & 0.004 & 0.000 & 0.937\\
\hspace{1em}$\bar{a}_{41}$ & -0.002 & 0.004 & 0.000 & 0.952 & 0.000 & 0.004 & 0.000 & 0.956\\
\hspace{1em}$\bar{a}_{42}$ & -0.005 & 0.005 & 0.000 & 0.921 & 0.000 & 0.005 & 0.000 & 0.944\\
\hspace{1em}$\bar{a}_{43}$ & 0.002 & 0.009 & 0.000 & 1.000 & 0.001 & 0.009 & 0.000 & 0.937\\
\hspace{1em}$\bar{a}_{44}$ & 0.004 & 0.011 & 0.000 & 0.996 & -0.001 & 0.011 & 0.000 & 0.925\\
\hspace{1em}$\bar{b}_{1}$ & 0.001 & 0.009 & 0.000 & 0.980 & -0.003 & 0.008 & 0.000 & 0.925\\
\hspace{1em}$\bar{b}_{2}$ & -0.001 & 0.004 & 0.000 & 1.000 & -0.034 & 0.004 & 0.001 & 0.000\\
\hspace{1em}$\bar{b}_{3}$ & -0.001 & 0.003 & 0.000 & 1.000 & -0.026 & 0.003 & 0.001 & 0.000\\
\hspace{1em}$\bar{b}_{4}$ & -0.001 & 0.003 & 0.000 & 1.000 & -0.025 & 0.003 & 0.001 & 0.000\\
\addlinespace[0.3em]
\multicolumn{9}{l}{\textbf{Random effects}}\\
\hspace{1em}$\sigma_{12}$ & 0.163 & 0.006 & 0.027 & 0.000 & - & - & - & -\\
\hspace{1em}$\sigma_{13}$ & 0.298 & 0.024 & 0.090 & 0.000 & - & - & - & -\\
\hspace{1em}$\sigma_{14}$ & 0.553 & 0.074 & 0.312 & 0.000 & - & - & - & -\\
\hspace{1em}$\sigma_{22}$ & 0.167 & 0.006 & 0.028 & 0.000 & - & - & - & -\\
\hspace{1em}$\sigma_{23}$ & 0.208 & 0.011 & 0.044 & 0.000 & - & - & - & -\\
\hspace{1em}$\sigma_{24}$ & 0.496 & 0.059 & 0.249 & 0.000 & - & - & - & -\\
\hspace{1em}$\sigma_{32}$ & 0.278 & 0.023 & 0.078 & 0.000 & - & - & - & -\\
\hspace{1em}$\sigma_{33}$ & 0.253 & 0.020 & 0.065 & 0.000 & - & - & - & -\\
\hspace{1em}$\sigma_{34}$ & 0.366 & 0.036 & 0.135 & 0.000 & - & - & - & -\\
\hspace{1em}$\sigma_{42}$ & 0.638 & 0.103 & 0.417 & 0.000 & - & - & - & -\\
\hspace{1em}$\sigma_{43}$ & 0.499 & 0.059 & 0.253 & 0.000 & - & - & - & -\\
\hspace{1em}$\sigma_{44}$ & 0.434 & 0.048 & 0.191 & 0.000 & - & - & - & -\\
\hspace{1em}$\tau_{1}$ & 0.002 & 0.000 & 0.000 & 0.000 & - & - & - & -\\
\hspace{1em}$\tau_{2}$ & 0.001 & 0.000 & 0.000 & 0.000 & - & - & - & -\\
\hspace{1em}$\tau_{3}$ & 0.002 & 0.000 & 0.000 & 0.000 & - & - & - & -\\
\hspace{1em}$\tau_{4}$ & 0.003 & 0.000 & 0.000 & 0.000 & - & - & - & -\\
\bottomrule
\end{tabular}
\label{table:supp_1}
\end{table*}

\begin{table*}

\caption{Performance metrics for the two models under conditions of between heterogeneity in the transitions}
\centering
\fontsize{10}{12}\selectfont
\begin{tabular}[t]{ccccccccc}
\toprule
\multicolumn{1}{c}{ } & \multicolumn{4}{c}{MHMM} & \multicolumn{4}{c}{HMM} \\
\cmidrule(l{3pt}r{3pt}){2-5} \cmidrule(l{3pt}r{3pt}){6-9}
parameter & bias & emp. SE & MSE & coverage & bias & emp. SE & MSE & coverage\\
\midrule
\addlinespace[0.3em]
\multicolumn{9}{l}{\textbf{Fixed effects}}\\
\hspace{1em}$\bar{a}_{11}$ & 0.002 & 0.015 & 0.000 & 0.944 & -0.008 & 0.015 & 0.000 & 0.336\\
\hspace{1em}$\bar{a}_{12}$ & -0.002 & 0.015 & 0.000 & 0.956 & -0.001 & 0.014 & 0.000 & 0.312\\
\hspace{1em}$\bar{a}_{13}$ & 0.000 & 0.003 & 0.000 & 0.968 & 0.007 & 0.004 & 0.000 & 0.152\\
\hspace{1em}$\bar{a}_{14}$ & 0.000 & 0.001 & 0.000 & 0.952 & 0.002 & 0.001 & 0.000 & 0.288\\
\hspace{1em}$\bar{a}_{21}$ & 0.002 & 0.022 & 0.000 & 0.940 & -0.036 & 0.019 & 0.002 & 0.084\\
\hspace{1em}$\bar{a}_{22}$ & -0.003 & 0.035 & 0.001 & 0.948 & 0.032 & 0.032 & 0.002 & 0.144\\
\hspace{1em}$\bar{a}_{23}$ & 0.000 & 0.019 & 0.000 & 0.952 & 0.000 & 0.019 & 0.000 & 0.328\\
\hspace{1em}$\bar{a}_{24}$ & 0.000 & 0.002 & 0.000 & 0.956 & 0.003 & 0.003 & 0.000 & 0.460\\
\hspace{1em}$\bar{a}_{31}$ & -0.001 & 0.008 & 0.000 & 0.924 & -0.012 & 0.007 & 0.000 & 0.248\\
\hspace{1em}$\bar{a}_{32}$ & 0.000 & 0.034 & 0.001 & 0.952 & -0.041 & 0.031 & 0.003 & 0.128\\
\hspace{1em}$\bar{a}_{33}$ & 0.000 & 0.039 & 0.002 & 0.948 & 0.048 & 0.037 & 0.004 & 0.108\\
\hspace{1em}$\bar{a}_{34}$ & -0.001 & 0.011 & 0.000 & 0.932 & 0.005 & 0.012 & 0.000 & 0.448\\
\hspace{1em}$\bar{a}_{41}$ & 0.000 & 0.006 & 0.000 & 0.924 & -0.006 & 0.005 & 0.000 & 0.496\\
\hspace{1em}$\bar{a}_{42}$ & -0.001 & 0.009 & 0.000 & 0.952 & -0.002 & 0.008 & 0.000 & 0.572\\
\hspace{1em}$\bar{a}_{43}$ & -0.002 & 0.026 & 0.001 & 0.936 & -0.032 & 0.022 & 0.002 & 0.176\\
\hspace{1em}$\bar{a}_{44}$ & 0.002 & 0.033 & 0.001 & 0.924 & 0.040 & 0.029 & 0.002 & 0.192\\
\hspace{1em}$\bar{b}_{1}$ & 0.000 & 0.011 & 0.000 & 0.976 & -0.004 & 0.010 & 0.000 & 0.916\\
\hspace{1em}$\bar{b}_{2}$ & -0.001 & 0.004 & 0.000 & 1.000 & -0.035 & 0.005 & 0.001 & 0.000\\
\hspace{1em}$\bar{b}_{3}$ & -0.001 & 0.003 & 0.000 & 1.000 & -0.022 & 0.004 & 0.000 & 0.000\\
\hspace{1em}$\bar{b}_{4}$ & -0.001 & 0.003 & 0.000 & 1.000 & -0.016 & 0.003 & 0.000 & 0.000\\
\addlinespace[0.3em]
\multicolumn{9}{l}{\textbf{Random effects}}\\
\hspace{1em}$\sigma_{12}$ & 0.059 & 0.192 & 0.040 & 0.952 & - & - & - & -\\
\hspace{1em}$\sigma_{13}$ & 0.047 & 0.215 & 0.048 & 0.980 & - & - & - & -\\
\hspace{1em}$\sigma_{14}$ & 0.090 & 0.268 & 0.080 & 0.984 & - & - & - & -\\
\hspace{1em}$\sigma_{22}$ & 0.037 & 0.180 & 0.034 & 0.968 & - & - & - & -\\
\hspace{1em}$\sigma_{23}$ & 0.062 & 0.216 & 0.050 & 0.956 & - & - & - & -\\
\hspace{1em}$\sigma_{24}$ & 0.099 & 0.294 & 0.096 & 0.972 & - & - & - & -\\
\hspace{1em}$\sigma_{32}$ & 0.078 & 0.212 & 0.051 & 0.972 & - & - & - & -\\
\hspace{1em}$\sigma_{33}$ & 0.043 & 0.202 & 0.043 & 0.972 & - & - & - & -\\
\hspace{1em}$\sigma_{34}$ & 0.078 & 0.224 & 0.056 & 0.988 & - & - & - & -\\
\hspace{1em}$\sigma_{42}$ & 0.181 & 0.352 & 0.156 & 0.992 & - & - & - & -\\
\hspace{1em}$\sigma_{43}$ & 0.151 & 0.274 & 0.098 & 0.976 & - & - & - & -\\
\hspace{1em}$\sigma_{44}$ & 0.097 & 0.270 & 0.082 & 0.964 & - & - & - & -\\
\hspace{1em}$\tau_{1}$ & 0.002 & 0.000 & 0.000 & 0.000 & - & - & - & -\\
\hspace{1em}$\tau_{2}$ & 0.001 & 0.000 & 0.000 & 0.000 & - & - & - & -\\
\hspace{1em}$\tau_{3}$ & 0.002 & 0.000 & 0.000 & 0.000 & - & - & - & -\\
\hspace{1em}$\tau_{4}$ & 0.004 & 0.000 & 0.000 & 0.000 & - & - & - & -\\
\bottomrule
\end{tabular}
\label{table:supp_2}
\end{table*}

\begin{table*}

\caption{Performance metrics for the two models under conditions of between heterogeneity in the emissions}
\centering
\fontsize{10}{12}\selectfont
\begin{tabular}[t]{ccccccccc}
\toprule
\multicolumn{1}{c}{ } & \multicolumn{4}{c}{MHMM} & \multicolumn{4}{c}{HMM} \\
\cmidrule(l{3pt}r{3pt}){2-5} \cmidrule(l{3pt}r{3pt}){6-9}
parameter & bias & emp. SE & MSE & coverage & bias & emp. SE & MSE & coverage\\
\midrule
\addlinespace[0.3em]
\multicolumn{9}{l}{\textbf{Fixed effects}}\\
\hspace{1em}$\bar{a}_{11}$ & 0.022 & 0.013 & 0.001 & 0.448 & 0.008 & 0.010 & 0.000 & 0.356\\
\hspace{1em}$\bar{a}_{12}$ & -0.021 & 0.016 & 0.001 & 0.576 & -0.027 & 0.010 & 0.001 & 0.016\\
\hspace{1em}$\bar{a}_{13}$ & -0.001 & 0.005 & 0.000 & 0.836 & 0.015 & 0.009 & 0.000 & 0.048\\
\hspace{1em}$\bar{a}_{14}$ & 0.000 & 0.001 & 0.000 & 0.928 & 0.004 & 0.003 & 0.000 & 0.148\\
\hspace{1em}$\bar{a}_{21}$ & -0.006 & 0.010 & 0.000 & 0.992 & -0.009 & 0.017 & 0.000 & 0.460\\
\hspace{1em}$\bar{a}_{22}$ & 0.038 & 0.019 & 0.002 & 0.560 & 0.038 & 0.023 & 0.002 & 0.132\\
\hspace{1em}$\bar{a}_{23}$ & -0.036 & 0.020 & 0.002 & 0.460 & -0.045 & 0.014 & 0.002 & 0.008\\
\hspace{1em}$\bar{a}_{24}$ & 0.003 & 0.003 & 0.000 & 0.888 & 0.016 & 0.007 & 0.000 & 0.012\\
\hspace{1em}$\bar{a}_{31}$ & 0.013 & 0.012 & 0.000 & 0.780 & 0.062 & 0.015 & 0.004 & 0.004\\
\hspace{1em}$\bar{a}_{32}$ & -0.044 & 0.025 & 0.003 & 0.564 & -0.089 & 0.022 & 0.008 & 0.000\\
\hspace{1em}$\bar{a}_{33}$ & 0.042 & 0.021 & 0.002 & 0.624 & 0.036 & 0.022 & 0.002 & 0.192\\
\hspace{1em}$\bar{a}_{34}$ & -0.012 & 0.007 & 0.000 & 0.704 & -0.010 & 0.011 & 0.000 & 0.392\\
\hspace{1em}$\bar{a}_{41}$ & 0.005 & 0.006 & 0.000 & 0.900 & 0.030 & 0.013 & 0.001 & 0.032\\
\hspace{1em}$\bar{a}_{42}$ & 0.012 & 0.012 & 0.000 & 0.796 & 0.053 & 0.016 & 0.003 & 0.000\\
\hspace{1em}$\bar{a}_{43}$ & -0.039 & 0.018 & 0.002 & 0.432 & -0.035 & 0.019 & 0.002 & 0.168\\
\hspace{1em}$\bar{a}_{44}$ & 0.020 & 0.013 & 0.001 & 0.904 & -0.048 & 0.017 & 0.003 & 0.044\\
\hspace{1em}$\bar{b}_{1}$ & 0.000 & 0.129 & 0.017 & 0.948 & 0.355 & 0.147 & 0.148 & 0.004\\
\hspace{1em}$\bar{b}_{2}$ & -0.147 & 0.149 & 0.044 & 0.616 & 0.183 & 0.199 & 0.073 & 0.036\\
\hspace{1em}$\bar{b}_{3}$ & 0.031 & 0.106 & 0.012 & 0.744 & 0.245 & 0.190 & 0.096 & 0.020\\
\hspace{1em}$\bar{b}_{4}$ & 0.078 & 0.061 & 0.010 & 0.640 & 0.196 & 0.125 & 0.054 & 0.016\\
\addlinespace[0.3em]
\multicolumn{9}{l}{\textbf{Random effects}}\\
\hspace{1em}$\sigma_{12}$ & 0.498 & 0.296 & 0.335 & 0.000 & - & - & - & -\\
\hspace{1em}$\sigma_{13}$ & 1.266 & 0.364 & 1.735 & 0.000 & - & - & - & -\\
\hspace{1em}$\sigma_{14}$ & 1.125 & 0.407 & 1.430 & 0.000 & - & - & - & -\\
\hspace{1em}$\sigma_{22}$ & 0.260 & 0.053 & 0.070 & 0.000 & - & - & - & -\\
\hspace{1em}$\sigma_{23}$ & 1.122 & 0.517 & 1.525 & 0.000 & - & - & - & -\\
\hspace{1em}$\sigma_{24}$ & 1.798 & 0.448 & 3.434 & 0.000 & - & - & - & -\\
\hspace{1em}$\sigma_{32}$ & 0.972 & 0.573 & 1.272 & 0.000 & - & - & - & -\\
\hspace{1em}$\sigma_{33}$ & 0.492 & 0.098 & 0.252 & 0.000 & - & - & - & -\\
\hspace{1em}$\sigma_{34}$ & 1.126 & 0.353 & 1.391 & 0.000 & - & - & - & -\\
\hspace{1em}$\sigma_{42}$ & 0.974 & 0.394 & 1.102 & 0.000 & - & - & - & -\\
\hspace{1em}$\sigma_{43}$ & 1.298 & 0.518 & 1.951 & 0.000 & - & - & - & -\\
\hspace{1em}$\sigma_{44}$ & 0.544 & 0.133 & 0.314 & 0.000 & - & - & - & -\\
\hspace{1em}$\tau_{1}$ & -0.080 & 0.160 & 0.032 & 0.960 & - & - & - & -\\
\hspace{1em}$\tau_{2}$ & -0.238 & 0.120 & 0.071 & 0.488 & - & - & - & -\\
\hspace{1em}$\tau_{3}$ & -0.309 & 0.038 & 0.097 & 0.000 & - & - & - & -\\
\hspace{1em}$\tau_{4}$ & -0.069 & 0.028 & 0.005 & 0.456 & - & - & - & -\\
\bottomrule
\end{tabular}
\label{table:supp_3}
\end{table*}

\begin{table*}

\caption{Performance metrics for the two models under conditions of between heterogeneity in transitions and emissions}
\centering
\fontsize{10}{12}\selectfont
\begin{tabular}[t]{ccccccccc}
\toprule
\multicolumn{1}{c}{ } & \multicolumn{4}{c}{MHMM} & \multicolumn{4}{c}{HMM} \\
\cmidrule(l{3pt}r{3pt}){2-5} \cmidrule(l{3pt}r{3pt}){6-9}
parameter & bias & emp. SE & MSE & coverage & bias & emp. SE & MSE & coverage\\
\midrule
\addlinespace[0.3em]
\multicolumn{9}{l}{\textbf{Fixed effects}}\\
\hspace{1em}$\bar{a}_{11}$ & 0.032 & 0.019 & 0.001 & 0.563 & 0.005 & 0.017 & 0.000 & 0.298\\
\hspace{1em}$\bar{a}_{12}$ & -0.030 & 0.021 & 0.001 & 0.619 & -0.025 & 0.015 & 0.001 & 0.119\\
\hspace{1em}$\bar{a}_{13}$ & -0.003 & 0.005 & 0.000 & 0.849 & 0.015 & 0.010 & 0.000 & 0.087\\
\hspace{1em}$\bar{a}_{14}$ & 0.000 & 0.001 & 0.000 & 0.933 & 0.005 & 0.004 & 0.000 & 0.119\\
\hspace{1em}$\bar{a}_{21}$ & -0.008 & 0.026 & 0.001 & 0.940 & -0.037 & 0.022 & 0.002 & 0.071\\
\hspace{1em}$\bar{a}_{22}$ & 0.050 & 0.040 & 0.004 & 0.726 & 0.071 & 0.033 & 0.006 & 0.063\\
\hspace{1em}$\bar{a}_{23}$ & -0.047 & 0.026 & 0.003 & 0.480 & -0.050 & 0.020 & 0.003 & 0.028\\
\hspace{1em}$\bar{a}_{24}$ & 0.004 & 0.004 & 0.000 & 0.873 & 0.016 & 0.009 & 0.000 & 0.024\\
\hspace{1em}$\bar{a}_{31}$ & 0.015 & 0.014 & 0.000 & 0.825 & 0.033 & 0.019 & 0.001 & 0.079\\
\hspace{1em}$\bar{a}_{32}$ & -0.064 & 0.043 & 0.006 & 0.631 & -0.109 & 0.026 & 0.013 & 0.000\\
\hspace{1em}$\bar{a}_{33}$ & 0.057 & 0.045 & 0.005 & 0.734 & 0.090 & 0.039 & 0.010 & 0.028\\
\hspace{1em}$\bar{a}_{34}$ & -0.011 & 0.012 & 0.000 & 0.817 & -0.014 & 0.016 & 0.000 & 0.262\\
\hspace{1em}$\bar{a}_{41}$ & 0.006 & 0.007 & 0.000 & 0.905 & 0.018 & 0.015 & 0.001 & 0.250\\
\hspace{1em}$\bar{a}_{42}$ & 0.011 & 0.017 & 0.000 & 0.837 & 0.038 & 0.020 & 0.002 & 0.056\\
\hspace{1em}$\bar{a}_{43}$ & -0.054 & 0.027 & 0.004 & 0.429 & -0.057 & 0.024 & 0.004 & 0.044\\
\hspace{1em}$\bar{a}_{44}$ & 0.034 & 0.028 & 0.002 & 0.829 & 0.000 & 0.036 & 0.001 & 0.349\\
\hspace{1em}$\bar{b}_{1}$ & 0.005 & 0.131 & 0.017 & 0.952 & 0.484 & 0.211 & 0.279 & 0.004\\
\hspace{1em}$\bar{b}_{2}$ & -0.103 & 0.121 & 0.025 & 0.754 & 0.322 & 0.262 & 0.172 & 0.020\\
\hspace{1em}$\bar{b}_{3}$ & 0.060 & 0.088 & 0.011 & 0.694 & 0.384 & 0.259 & 0.214 & 0.000\\
\hspace{1em}$\bar{b}_{4}$ & 0.088 & 0.066 & 0.012 & 0.567 & 0.283 & 0.184 & 0.114 & 0.016\\
\addlinespace[0.3em]
\multicolumn{9}{l}{\textbf{Random effects}}\\
\hspace{1em}$\sigma_{12}$ & 0.685 & 0.570 & 0.793 & 0.484 & - & - & - & -\\
\hspace{1em}$\sigma_{13}$ & 1.550 & 0.763 & 2.981 & 0.063 & - & - & - & -\\
\hspace{1em}$\sigma_{14}$ & 0.904 & 0.675 & 1.271 & 0.528 & - & - & - & -\\
\hspace{1em}$\sigma_{22}$ & 0.152 & 0.240 & 0.081 & 0.933 & - & - & - & -\\
\hspace{1em}$\sigma_{23}$ & 1.562 & 0.824 & 3.115 & 0.087 & - & - & - & -\\
\hspace{1em}$\sigma_{24}$ & 1.680 & 0.822 & 3.497 & 0.127 & - & - & - & -\\
\hspace{1em}$\sigma_{32}$ & 1.529 & 1.222 & 3.826 & 0.306 & - & - & - & -\\
\hspace{1em}$\sigma_{33}$ & 0.368 & 0.355 & 0.261 & 0.742 & - & - & - & -\\
\hspace{1em}$\sigma_{34}$ & 1.075 & 0.623 & 1.542 & 0.341 & - & - & - & -\\
\hspace{1em}$\sigma_{42}$ & 0.901 & 0.916 & 1.648 & 0.659 & - & - & - & -\\
\hspace{1em}$\sigma_{43}$ & 1.648 & 1.096 & 3.913 & 0.262 & - & - & - & -\\
\hspace{1em}$\sigma_{44}$ & 0.249 & 0.364 & 0.194 & 0.877 & - & - & - & -\\
\hspace{1em}$\tau_{1}$ & -0.090 & 0.159 & 0.033 & 0.921 & - & - & - & -\\
\hspace{1em}$\tau_{2}$ & -0.330 & 0.084 & 0.116 & 0.147 & - & - & - & -\\
\hspace{1em}$\tau_{3}$ & -0.326 & 0.036 & 0.108 & 0.000 & - & - & - & -\\
\hspace{1em}$\tau_{4}$ & -0.074 & 0.027 & 0.006 & 0.333 & - & - & - & -\\
\bottomrule
\end{tabular}
\label{table:supp_4}
\end{table*}


\begin{figure*}[h]
\centering
\includegraphics[width=0.7\linewidth]{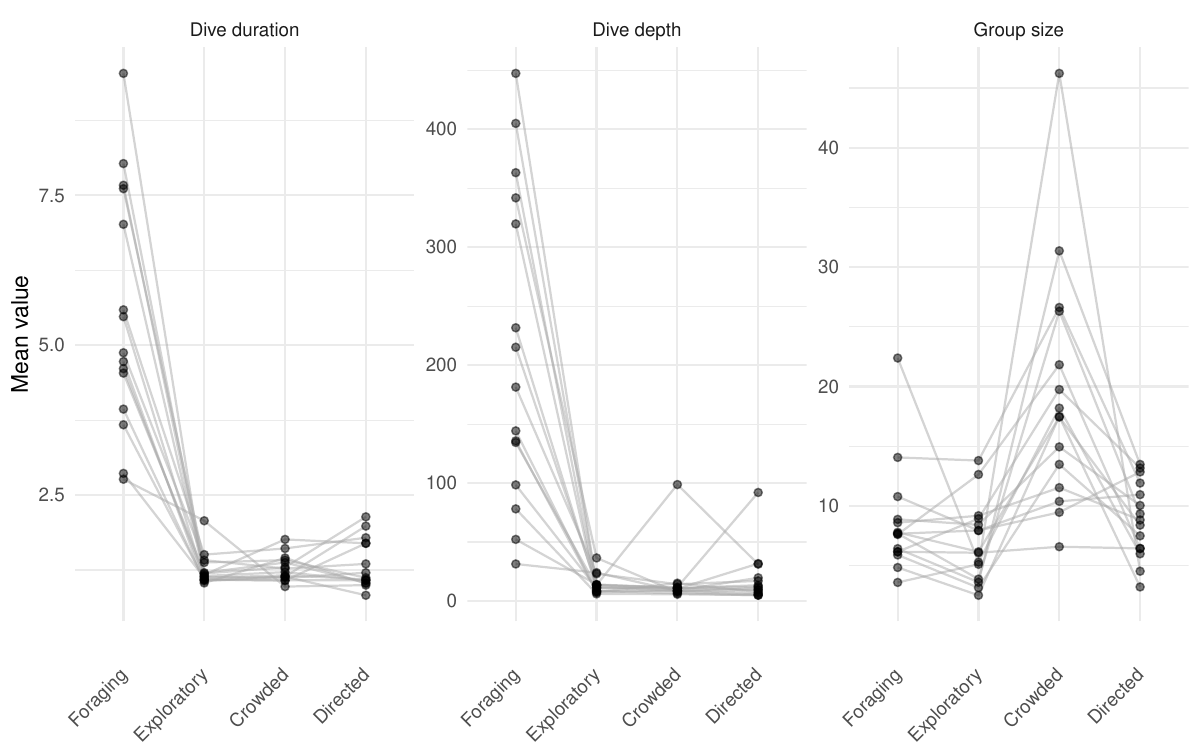}
\caption{The emission means are relatively stable across whales, which indicates that the behavioral states have similar representations across the whales. The state-specific emission means for each whale (\textit{maximum a posteriori} estimates) have been linked to facilitate the comparison. Parallel lines indicate similar emission pattern across states for different whales.}
\label{fig:whale_emiss_linked}
\end{figure*}


\begin{figure*}[h]
\centering
\includegraphics[width=0.5\linewidth]{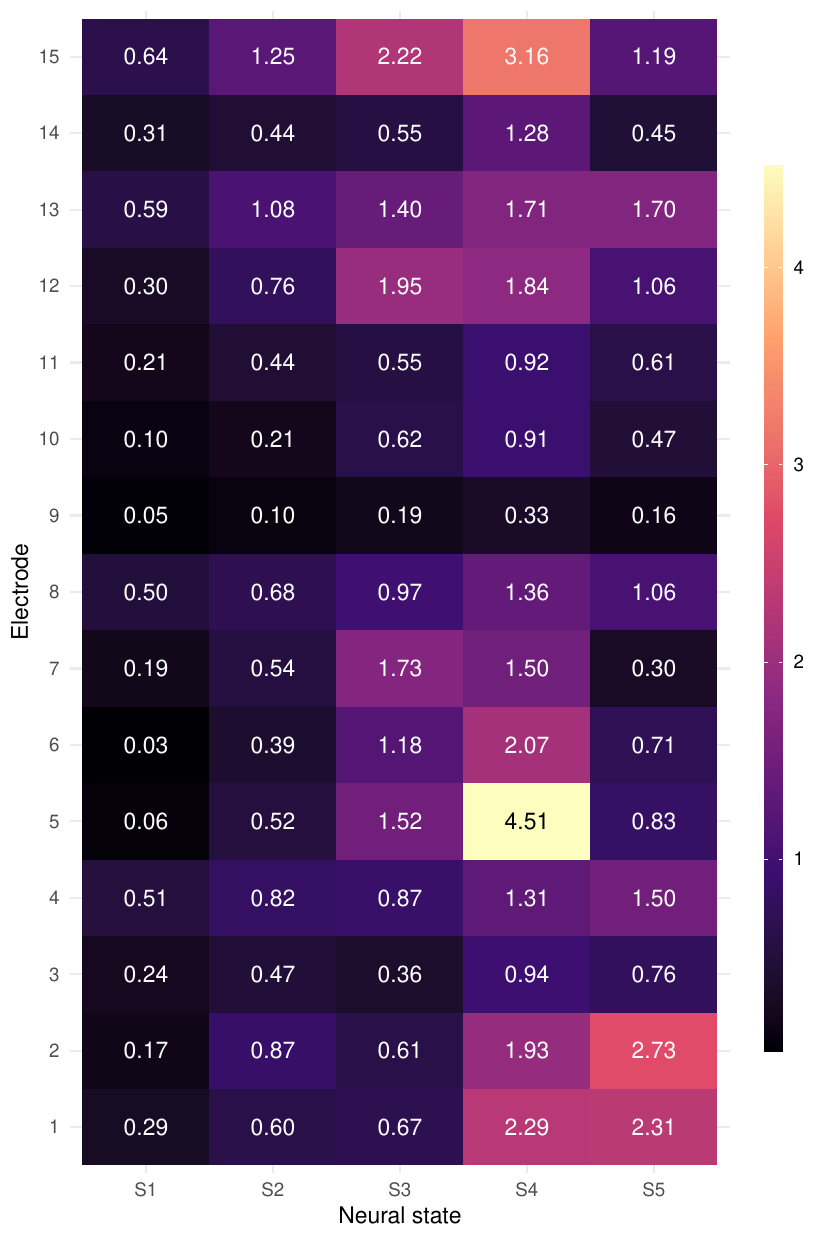}
\caption{Group-level Poisson emission means (\textit{maximum a posteriori}) by electrode and state, displayed in the natural scale as $exp(\bar{b}_{ki})$ to facilitate comprehension.}
\label{fig:monkey_group_emiss}
\end{figure*}

\begin{figure*}[h]
\centering
\includegraphics[width=0.5\linewidth]{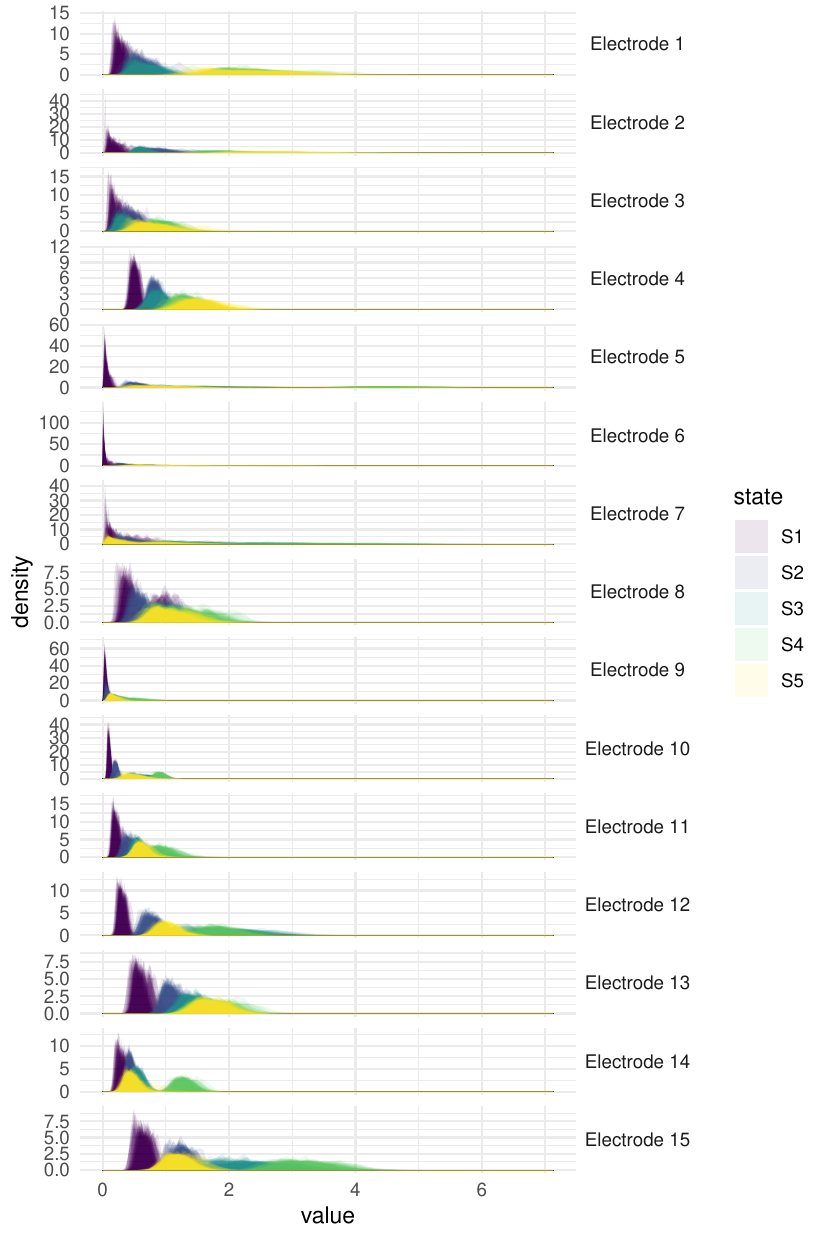}
\caption{The spread in the trial-specific posterior samples of the Poisson emission parameters $b_{ki}$ shows the variability in the emission distribution parameters between trials.}
\label{fig:monkey_trial_emiss}
\end{figure*}

\end{document}